\documentclass[letterpaper,pdftex,rmp,twocolumn,amsmath,amssymb,groupedaddress,floatfix]{revtex4}

\usepackage{graphicx}
\usepackage{natbib}
\usepackage{textcomp}
\usepackage[normalem]{ulem}
\usepackage{makecell}
\usepackage{float}
\usepackage{setspace}
\usepackage[a]{esvect}
\usepackage[font=sf,size=footnotesize,justification=RaggedRight,singlelinecheck=false]{caption}
\usepackage[sf,bf,small]{titlesec}
\titlespacing*{\section}{0pt}{1em}{0em}
\usepackage{xcolor}
\usepackage{placeins}

\usepackage{mathrsfs}
\usepackage{amsmath,amssymb}
\usepackage{bm}% bold math

\usepackage{float}

\definecolor{darkgray}{rgb}{0.25,0.25,0.25}
\definecolor{darkred}{rgb}{0.89,0.10,0.11}
\definecolor{darkblue}{rgb}{0.12,0.39,0.62}
\usepackage{url}

\usepackage[pdftex,breaklinks=true,colorlinks=true,citecolor=black,linkcolor=black,menucolor=black,urlcolor=darkblue,pdfborder={1 0 0}]{hyperref}
\hypersetup{pdftitle={Diffusion on networked systems is a question of time or structure},pdfauthor={J.-C. Delvenne, R. Lambiotte and L.E.C. Rocha (2015)}}
\bibpunct{()}{)}{,}{s}{,}{,}

\usepackage{multirow}
\usepackage{booktabs}
\usepackage{array}
\usepackage{color,soul}

\begin{document}
\makeatletter
\renewcommand\@biblabel[1]{#1.}
\makeatother

\newcommand{\dif}{\mathrm{d}}

\renewcommand{\figurename}{Figure}
\renewcommand{\thefigure}{\arabic{figure}}
\renewcommand{\tablename}{Table}
\renewcommand{\thetable}{\arabic{table}}
\renewcommand{\refname}{\large References}

\addtolength{\textheight}{1cm}
\addtolength{\textwidth}{1cm}
\addtolength{\hoffset}{-0.5cm}

\setlength{\belowcaptionskip}{1ex}
\setlength{\textfloatsep}{2ex}
\setlength{\dbltextfloatsep}{2ex}

\hyphenation{page-rank}

\title{Diffusion on networked systems is a question of time or structure}

%\date{\today}

\author{Jean-Charles Delvenne}
\email{Jean-Charles.Delvenne@uclouvain.be}
\affiliation{ICTEAM and CORE, University of Louvain, 4 Avenue Lema\^{i}tre, B-1348 Louvain-la-Neuve, Belgium}

\author{Renaud Lambiotte}
\affiliation{Department of Mathematics and naXys, Universit\'e de Namur, 8 Rempart de la Vierge, B-5000 Namur, Belgium}

\author{Luis E C Rocha}
\affiliation{
Department of Mathematics and naXys, Universit\'e de Namur, 8 Rempart de la Vierge, B-5000 Namur, Belgium \\
Department of Public Health Sciences, Karolinska Institutet, 18A Tomtebodav\"agen, S-17177 Stockholm, Sweden}

\begin{abstract}
Network science investigates the architecture of complex systems to understand their functional and dynamical properties. Structural patterns such as communities shape diffusive processes on networks. However, these results hold under the strong assumption that networks are static entities where temporal aspects can be neglected. Here we propose a generalised formalism for linear dynamics on complex networks, able to incorporate statistical properties of the timings at which events occur. We show that the diffusion dynamics is affected by the network community structure and by the temporal properties of waiting times between events. We identify the main mechanism --- network structure, burstiness or fat-tails of waiting times --- determining the relaxation times of stochastic processes on temporal networks, in the absence of temporal-structure correlations. We identify situations when fine-scale structure can be discarded from the description of the dynamics or, conversely, when a fully detailed model is required due to temporal heterogeneities.
\end{abstract}

\maketitle

\noindent 

The relation between network structure and dynamics has attracted the attention of researchers from different disciplines over the years~\cite{Moody09, Newman10, Bansal10, Barrat12, Vespignani12}. These works are rooted in the observation that, in contexts as diverse as the Internet, society, and biology, networks tend to possess complex patterns of connectivity, with a significant level of heterogeneity~\cite{Newman10}. In addition, a broad range of real-world dynamical processes, from information to virus spreading, is akin to diffusion. If the effect of structure, such as communities or degree heterogeneity, on diffusive processes is now well-known~\cite{Lovasz1993, Chung96}, the impact of the temporal properties of individual nodes is poorly understood~\cite{Bansal10, Holme12}. Yet, empirical evidence indicates that real-world networked systems are often characterised by complex temporal patterns of activity, including a fat-tailed distribution of times~\cite{Eckmann04, Holme04, Barabasi05, Rocha10, Isella11, Starnini12, Haerter12, Vanhems13}, correlations between events~\cite{Scholtes2014, Rosvall2014} and non-stationarity~\cite{Rocha13, Holme13, Horvath14}. This is in remarkable contrast to a vast majority of mathematical models, which assume homogeneous interaction dynamics on networks~\cite{Malmgren08, Bansal10, Barrat12, Perra2012}. In this work, we focus on the impact of structure and the waiting time distribution on dynamics, and set aside any other temporal properties. This subject has triggered intense theoretical work in recent years, for example in relation to anomalous diffusion, but predominantly on lattice-like, annealed~\cite{Klafter11} and random structures~\cite{Vazquez07, Iribarren09, Min11, Perra2012, Rocha13, Jo14}. These limitations leave a fundamental question open, with important applications for the modelling of temporal networks: what are the effects of complex patterns, simultaneously in time and in structure, on the approach to equilibrium of diffusion processes?

\begin{figure*}[htb]
\centering
\includegraphics[scale=1.0]{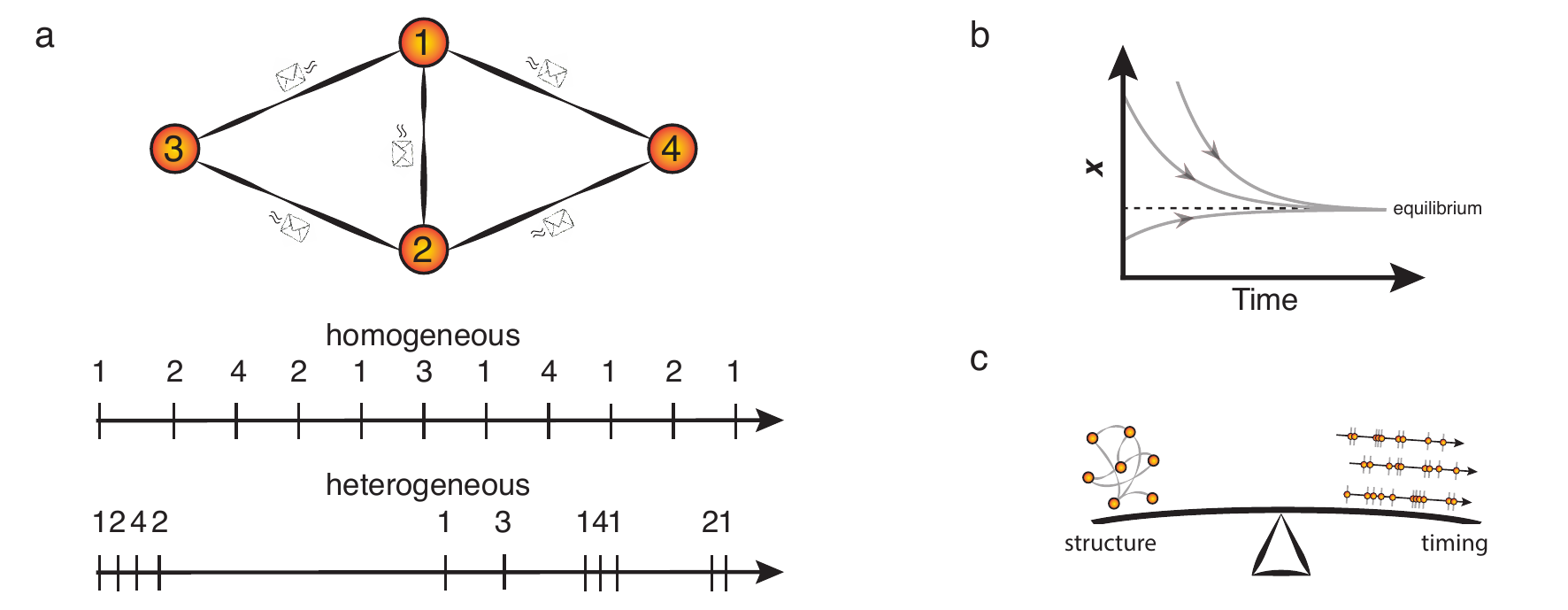}
\caption{\textbf{Diffusion on temporal networks.} (a) A random walk process can be illustrated by a letter (or banknote, etc.) being randomly passed from neighbour to neighbour on a social network. Temporal patterns of waiting times between arrival and departure of the letter may be homogeneous (for instance discrete or exponentially distributed times), or heterogeneous (for instance bursts). (b) The relaxation time measures the characteristic time to reach equilibrium from any starting condition. (c) The competition between structure and temporal patterns regulates the relaxation time, or mixing time, of stochastic processes.}
\label{fig1}
\end{figure*}

To comprehend the interplay between temporal and structural patterns, we focus on a broad class of linear multi-agent systems describing $N$ interacting nodes, defined by 
\begin{equation}
D \bm{\mathrm{x}}=L  \bm{\mathrm{x}}
\label{eq:DxLx}
\end{equation}
where $x_i$, the $i$th component of $\bm{\mathrm{x}}$, represents the observed state of node $i$. The $N \times N$ real matrix $L$ encodes the mutual influences in the network, with non-zero entries indicating the presence of a link. $D$ is either $\dif /\dif t$, the delay $Dx_i(t)=x_i(t+1)$, or any other causal operator acting linearly on the trajectory of each entry $x_i(t)$. This equation couples network structure (represented by $L$) and time evolution (represented by $D$) by describing a system where every node $i$ has a state $x_i(t)=F u_i(t)$, where $u_i(t)=\sum_j L_{ij} x_j(t)$ represents the input, or influence of the neighbouring states on node $i$. The operator $F$ is the so-called transfer function~\cite{astrom2010feedback,Fax04}, defined as the inverse of $D$ (See Methods).

A classic example is heat diffusion on networks, where every node has a temperature $x_i$ evolving according to the Fourier Law
\begin{equation}
 \mu \dfrac{\dif \bm{\mathrm{x}}}{\dif t}=L \bm{\mathrm{x}},
\label{eq:diffusion}
\end{equation}
where $\mu$ is the characteristic time of the dynamics and $L$ is a Laplacian of the network. The same set of equations can represent, possibly up to a change of variables, a basic model for the evolution of people's opinions~\cite{Blondel2005}, robots' positions in the physical space~\cite{Fax04,ali}, approach to synchronisation~\cite{Strogatz2000,dorfler2012synchronization} or the dynamics of a continuous-time random walker~\cite{yin2012continuous}---our main example from now. In any case, the dynamical properties of the system are described by the spectral properties of the coupling matrix. The constraints imposed by the conservation of probability imply that the Laplacian dynamics is characterised by a stationary state, associated to the dominant eigenvector of $L$, which we will assume to be unique, as is the case in a large class of systems, for example strongly connected networks. A key quantity is thus the second dominant eigenvalue, also called the spectral gap~\cite{Chung96}, which provides us with the relaxation time to stationarity, usually called mixing time~\cite{Levinmixing} for stochastic diffusion processes. The spectral gap determines the speed of convergence to the stationary state, and measures the effective size of the system in terms of dynamics. The spectral gap is also related to important structural and dynamical properties of the system, such as the existence of bottlenecks and communities in the underlying network \cite{Lovasz1993,Fax04}.

In this paper, we generalise the concept of spectral gap and of mixing time to random processes with  general causal operator $D$, and focus in detail on operators with long-term memory, naturally emerging in diffusion with bursty dynamics.  After showing connections between the theory of random walks and that of multi-agents systems, we identify  the temporal and structural mechanisms driving the asymptotic dynamics of the system, and provide examples when each mechanism prevails. By doing so, we show that the form of the temporal operator $D$ may either slow-down or accelerate mixing as compared to the differential operator equation (\ref{eq:diffusion}). The results are further exploited to assess the possibility of coarse-graining the dynamics based on the network community structure, and tested using numerical simulations on synthetic temporal networks calibrated with empirical data.

\begin{figure}[htb]
\centering
\includegraphics[scale=0.5]{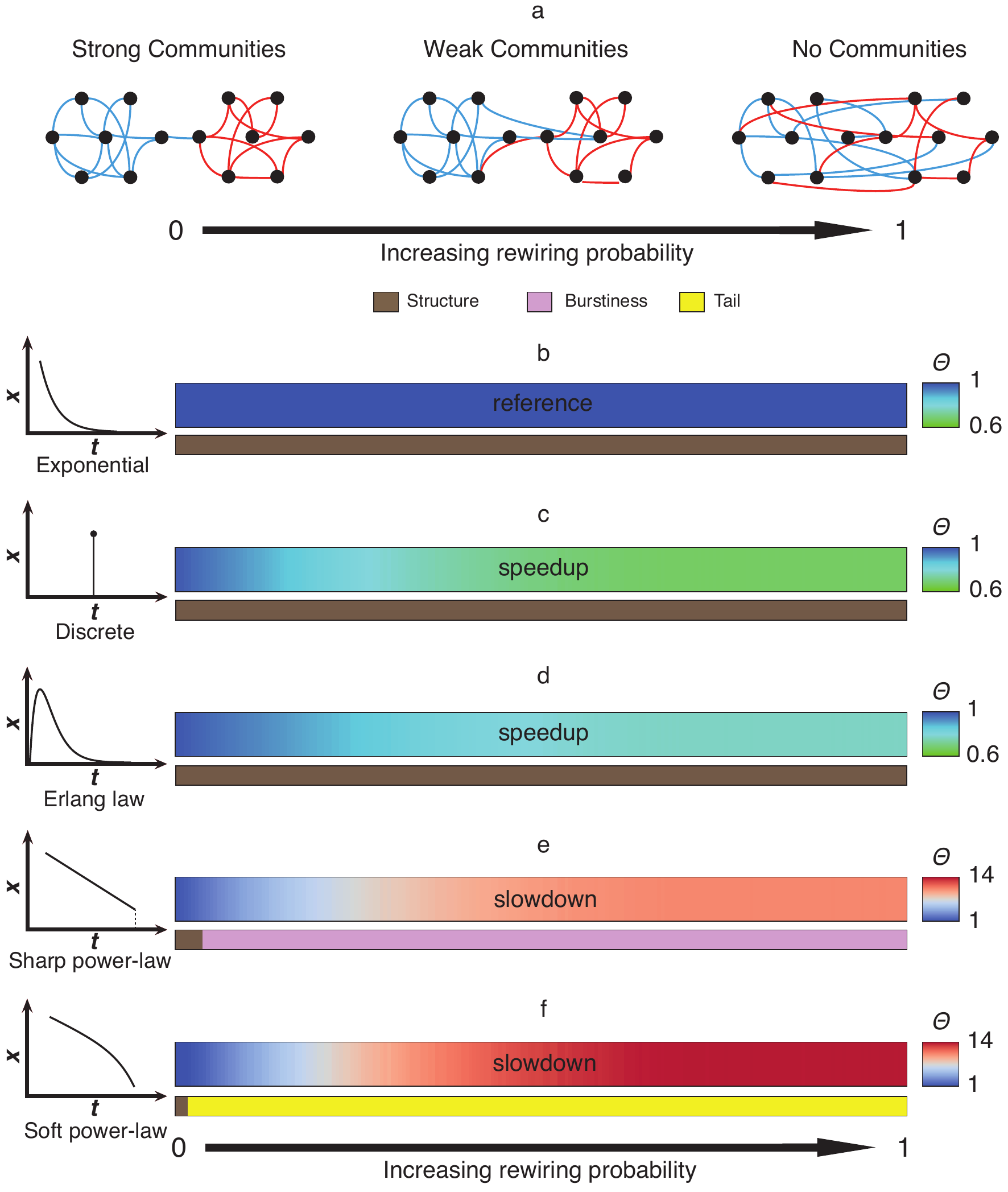}
\caption{\textbf{Network structure or waiting times regulate mixing times.} We illustrate the findings simple network composed of two communities, where the community structure is modulated by rewiring edges starting from two separated random graphs, as shown in a).  Each panel (b-f) shows (on the top bar) for each value of the rewiring probability the slowdown factor $\Theta$, ratio of the exact mixing time for a given waiting time distribution and for the exponential case (Poisson process). Each panel corresponds to a different waiting time distribution and the colorbar scale on the right represents the values of $\Theta$ for the respective colormap. $\Theta$ close to one indicates that non-trivial waiting time patterns can be neglected in the determination of the mixing time. On the other hand, a high $\Theta$ indicates that the details of the network structure are overshadowed by the temporal dynamics. b) shows the reference case, that is, the exponential waiting times. We observe a speedup ($\Theta<1$) for c) Dirac (discrete) and d) Erlang distributions and a slowdown ($\Theta>1$) for e) a power-law with sharp cutoff and for f) a power-law with soft cutoff. From left to right, we show how this factor depends on the network structure by progressively removing the community structure (increasing $\epsilon$). The bottom bar of each panel shows, according to equation~(\ref{eq:taumodetailmix}), which is the dominant factor, either structure, burstiness, or tail, regulating the relaxation time for each model of waiting times. See Methods for details on the synthetic networks and waiting time distributions.}
\label{fig2}
\end{figure}

\section*{Results}

\noindent\textbf{Random walks with arbitrary waiting times.}
The generalised dynamics of the random walker, illustrated in Fig.~\ref{fig1}a, is defined as follows. Assume that a walker arrives at node $i$ and hops at a time between $\Delta t$ and $\Delta t+\delta$ after its arrival. 
A walker arriving at a node $i$ jumps towards a neighbouring node within a time interval $[\Delta t, \Delta t + \delta]$ with probability $\rho(\Delta t) \delta$ (for small $\delta$). In line with a standard discrete time random walk process, the jump is directed towards a neighbour $j$ with probability $P_{ij}$. The probability density function $\rho(\Delta t)$ is called the waiting time distribution of the walker. At each hop, the waiting time $\Delta t$ is reset to zero, and consecutive waiting times are independent. The evolution of probability of presence of a stochastic process in each state is ruled by the so-called master equation, or Kolmogorov forward equation, well known for this family of generalised walks~\cite{montroll1965random, Klafter11, Hoffmann12}. We prefer to adopt here the equivalent, dual, viewpoint of Kolmogorov backward equation~\cite{kolmogoroff1931analytischen}, which belongs to the class of processes defined by equation (\ref{eq:DxLx}) and can be analysed using the toolbox typical to multi-agent systems~\cite{Fax04,ali}, such as  the transfer function formalism and  eigenmode decomposition.

We assign a fixed real-valued observable $x_{\textrm{obs,}i}$ to every node $i$, and consider the value observed at time $t$ by a walker starting initially from node $j$. This value is a random variable, with an expected value  denoted by $x_{j}(t)$, taking initial value $x_j(0)=x_{\textrm{obs,}j}$. The random walk is ergodic and mixing on a strongly connected aperiodic network (aperiodicity is only required for the discrete-time, when $\rho(\Delta t)$ is a Dirac distribution). In this case,  $\bm{\mathrm{x}}(t)$ describes a consensus dynamics, meaning that the individual values asymptotically converge to one another, $x_i(\infty)=x_j(\infty)$. If we choose the observable $x_{\textrm{obs,}j}=0$ at all nodes $j$, except $1$ at node $k$, then $x_{i}(t)$ is precisely the probability for the random walker starting at $i$ to be found at $k$ at time $t$. Therefore the Kolmogorov backward equation, a consensus dynamics, embeds in particular the evolution of the probability of presence on nodes.

The walker, assumed to have just hopped at time zero and finding itself at node $i$ (another origin of time would not change the asymptotic decay times, of main interest in this work), hops again at time $\Delta t$ with probability density $\rho(\Delta t)$, and moves towards a neighbour $j$ with probability $P_{ij}$. The expected value observed by the walker at time $t$ is $x_{\textrm{obs,}i}$ if it is still waiting for its first hop ($t< \Delta t$) and otherwise $\sum_j P_{ij} x_{j}(t-\Delta t)$, by induction on the number of hops. Therefore we obtain the vector equation
\begin{equation}
\bm{\mathrm{x}}(t)= \bm{\mathrm{x}_{\textrm{obs}}}  \int_t^{\infty}  \rho(\Delta t) \dif\Delta t + \int_0^t P \bm{\mathrm{x}}(t-\Delta t) \rho(\Delta t) \dif\Delta t,
\label{eq:M1}
\end{equation}
with the discrete transition matrix $P$ of entries $P_{ij}$. The convolution in time in the last term calls for a Laplace transform 
\begin{equation}\label{eq:modesv}
\bm{\mathrm{x}}\left(s\right) = \int_{0}^{\infty} \bm{\mathrm{x}}\left(t\right)e^{-st}\dif t.
\end{equation}
For simplicity, we use the same notations for functions in the time and in the Laplace domain, only distinguished by their variable, namely $t$ or $\Delta t$ for time, and $s$ for Laplace. This is justified as time and Laplace domain representations encode the same single physical object, for example a probability or an observable. The same holds for operator $D$, thought of as acting in the time domain (for instance $D=\dif/\dif t$) or the Laplace domain ($D=s$) according to the context.

The Laplace transform $\rho(s)$ is the moment generating function of the waiting time distribution $\rho(\Delta t)$, as it encodes the moment in its Taylor series $\rho(s)=1-\mu s + (\mu^2+\sigma^2)\frac{s^2}{2}- \cdots$, where $\mu$ is the expected waiting time (first moment), $\sigma^2$ is the variance and $\mu^2+\sigma^2$ is the second moment. Using the fact that convolution (respectively, integration from $0$ to $t$) in the time domain corresponds to the usual product (respectively, division by $s$) in the Laplace domain~\cite{dyke1999introduction}, equation~(\ref{eq:M1}) reduces to
 \begin{equation}
\bm{\mathrm{x}}(s)=   \frac{1-\rho(s)}{s} \bm{\mathrm{x}_{\textrm{obs}}} + \rho(s) P \bm{\mathrm{x}}(s),
\label{eq:M2}
\end{equation}
or equivalently 
\begin{equation}
\left(\frac{1}{\rho(s)}-1\right)\bm{\mathrm{x}}(s)=  \left(\frac{1}{\rho(s)}-1\right) \frac{1}{s} \bm{\mathrm{x}}(t=0) + L \bm{\mathrm{x}}(s),
\label{eq:M3}
\end{equation}
where we have made the dependence on the initial condition explicit by using the relation  $\bm{\mathrm{x}_{\textrm{obs}}}=\bm{\mathrm{x}}(t=0)$ and where $L=P-I$ denotes the (normalised) Laplacian of the network. 
This is an instance of equation~(\ref{eq:DxLx}), which shows that an input-output relationship is often best expressed in the Laplace domain rather than in the temporal domain. In this case, the  transfer function $F(s)$ is defined by the algebraic relation $F^{-1}(s)=1/\rho(s) -1=D(s)$, up to the initial condition term, implicit in equation~(\ref{eq:DxLx}). See Methods for a derivation of equation~(\ref{eq:DxLx}) in a more general context. 

\vspace{0.5cm}
\noindent\textbf{From temporal networks to random walks.}
Diffusive processes often take place on temporal networks where individuals initiate from time to time short-lived contacts with their neighbours. A random walker can represent for example a letter or a banknote passing from hand to hand through first contact initiated by the current node. The formalism described above focuses on the statistical properties of the waiting times of a walker on a node~{\cite{Iribarren09, Starnini12}, and not of the inter-contact times (the times between two subsequent contacts from a given node to another), as often considered in the literature~\cite{Vazquez07, Min11, karsai2011small, Rocha11, Kivela12, Rocha13}. To illustrate this difference, let us consider an idealised scenario, where the network looks locally like a directed tree, in order to avoid indirect correlations due to cycles~\cite{Speidel}, where the inter-contact time $t$ between two contacts initiated by the same node is characterised by the same probability distribution $\rho_\textrm{contact}(t)$, and where activations on different edges are an independent random process. The corresponding waiting time distribution $\rho(\Delta t)$ for the random walker can be determined from $\rho_\textrm{contact}$\cite{kleinrock1975queueing, Lambiotte13}.

For example, the classic inspection paradox, or bus paradox, observes that the waiting time has a mean 

\begin{equation}
\mu=1/2 \mu_{\textrm{contact}} (1+\sigma^2_{\textrm{contact}}/\mu^2_{\textrm{contact}})
\label{eq:contact}
\end{equation}
which can be much higher than the average inter-contact time $\mu_\textrm{contact}$ in case of bursts. This fact has been used in the literature to deduce that burst contact statistics slow down diffusion in a complex network~\cite{karsai2011small}. Nevertheless it should not be confounded with the results presented in this paper, which will focus on the statistical properties of the waiting time distribution and identify, amongst others, that its variance plays a significant role on the asymptotic behaviour of the walker. In the above scenario, the variance of $\rho(\Delta t)$ depends on the third moment of the inter-contact time distribution $\rho_\textrm{contact}(t)$, and is associated to a mechanism distinct from that of the bus paradox (\ref{eq:contact}).

\vspace{0.5cm}
\noindent\textbf{Eigenmodes for heterogeneous temporal operators.} Equation~(\ref{eq:DxLx}) typically takes the form of an integro-differential equation. However, it simplifies into the differential equation~(\ref{eq:diffusion}) in the case of a memoryless random walker. Memoryless refers to  the case when the probability of hopping between $\Delta t$ and $\Delta t + \delta$, knowing that the walker has waited at least $\Delta t$, is independent of $\Delta t$. This leads to an exponential waiting time distribution~\cite{kleinrock1975queueing} $\rho(\Delta t)=e^{-\Delta t/\mu}/\mu$, in other words an unconditional probability $e^{-\Delta t/\mu} \delta/\mu$ of jumping between $\Delta t$ and $\Delta t + \delta$, for small $\delta$ and for mean waiting time $\mu$. In that case, we find $D(s)=1/\rho(s) -1=\mu s$ in the Laplace domain, indeed recovering equation~(\ref{eq:diffusion}) in the time domain. 
The differential equation can then be analysed by changing the variables $\bm{\mathrm{x}}$ to a linear combination of the eigenvectors $\bm{\mathrm{v}}_k$ of the Laplacian $L$, of eigenvalues $\lambda_0=0 > \lambda_1, \lambda_2, \ldots, \lambda_{N-1} \geq -2$ as follows~\cite{Fax04}: 
\begin{equation}
\bm{\mathrm{x}}(t)=  \sum_k z_k(t) \bm{\mathrm{v}}_k.\label{eq:eigenmodes}
\end{equation}
For simplicity we have supposed that the underlying network is undirected, connected and, in case of a discrete-time random walk, non-bipartite. Thus the eigenvalues are real and the stationary state is uniquely defined~\cite{Chung96}. Every $z_k(t)$, solution of $Dz_k(t)=\lambda_k z_k(t)$, is called an eigenfunction of the operator $D$, and here takes the form of a decaying exponential. The problem is thus solved by decoupling  structural and temporal variables, first by identifying structural eigenvectors ($\bm{\mathrm{v}}_k$), and then how they evolve in time ($z_k(t)$). The resulting fundamental solutions $z_k(t) \bm{\mathrm{v}}_k$ for equation~(\ref{eq:DxLx})  are called modes, or eigenmodes, of the system.

\begin{figure}[htb]
\centering
\includegraphics[scale=0.65]{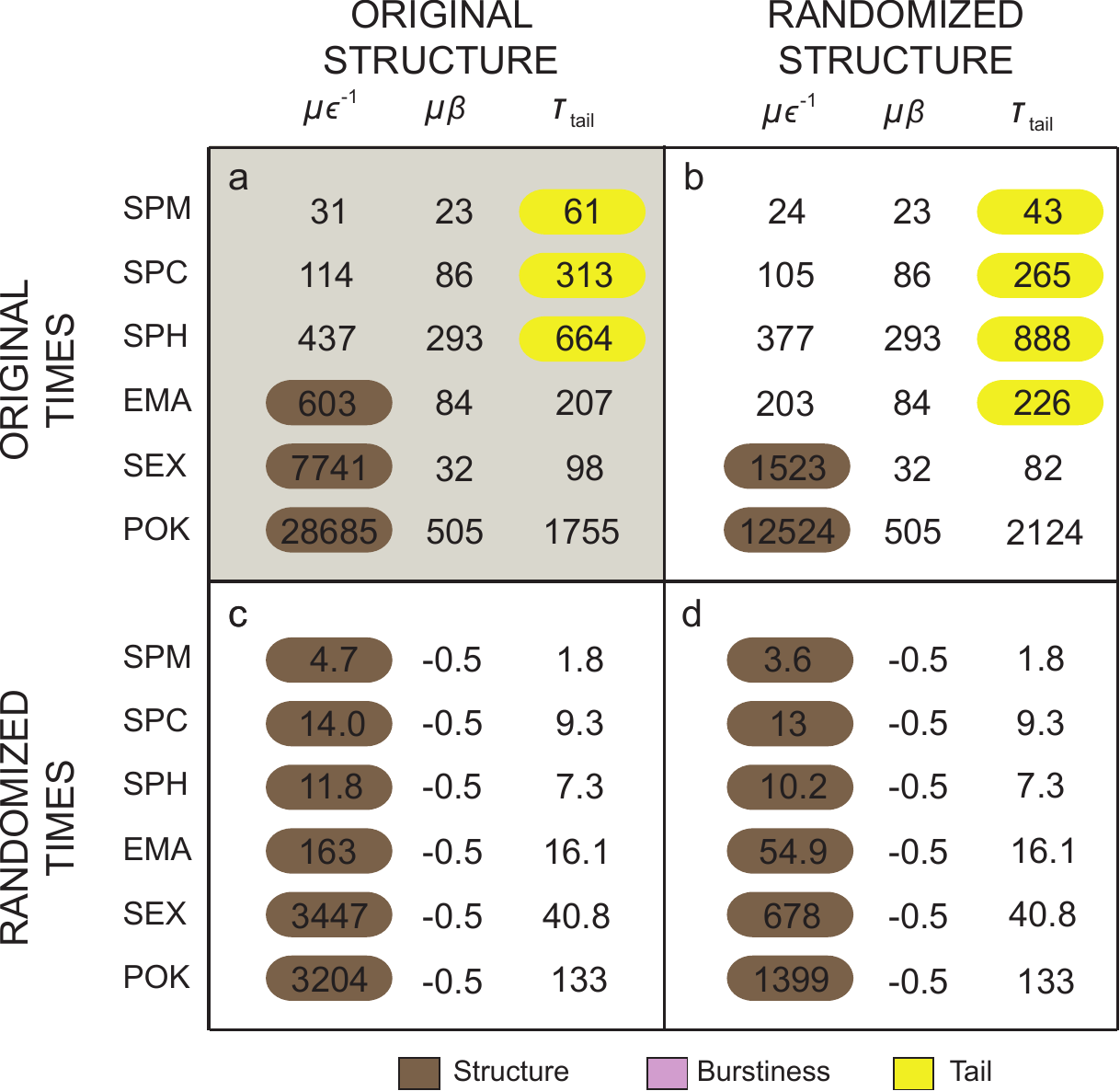}
\caption{\textbf{Dominating mechanisms on temporal networks.} (a) shows the regulating potential of each mechanism according to equation~(\ref{eq:taumodetailmix}), either structure ($\mu \epsilon^{-1}$), burstiness ($\mu \beta$), or tail ($\tau_{\textrm{tail}}$), for the synthetic networks based on the empirical network structures and on the waiting times obtained by simulating a random walk on these structures. Either structure (brown) or tail (yellow) dominates the mixing time, also when structure or contact times are randomized independently (b-d).}
\label{fig3arxiv}
\end{figure}

A similar analysis can be performed in the Laplace domain in the case of an arbitrary waiting time distribution $\rho(\Delta t)$, and thus when equation~(\ref{eq:DxLx}) has an arbitrary temporal operator $D$. In that case, the elementary solutions $z_k(t)$, associated to $z_k(s)$, need not be exponential functions, and are obtained as solutions of equation~(\ref{eq:M3}), where the Laplacian is reduced to its eigenvalue $\lambda_k$
\begin{equation}
z_k(s)=\frac{1}{1-\frac{\lambda_k}{1/\rho(s)-1}}\frac{z_k(t=0)}{s},
\label{eq:rwmodes}
\end{equation}
As before, any trajectory of the system can be expressed as a linear superposition of some or all the $N$ modes. 

\vspace{0.5cm}
\noindent\textbf{Characteristic decay times of eigenmodes.} Despite their non-exponential nature, a broad class of eigenfunctions (\ref{eq:rwmodes}) still have a characteristic time $\tau_k$ describing the asymptotic decay of $z_k(t)$ to equilibrium as $e^{-t/\tau_k}$ (Fig.~\ref{fig1}b). We find that the decay time can be accurately estimated by performing a small $s$ expansion in the Laplace domain (see Methods for the derivation and range of validity):
\begin{equation}
\tau_k \approx \mu (|\lambda_k|^{-1}+\beta),
\label{eq:taumode}
\end{equation}
where 
\begin{equation}
\beta =\frac{\sigma^2-\mu^2}{2 \mu^2}
\label{eq:beta}
\end{equation}
is a measure of the burstiness of the temporal process based on the first and second moments of its distribution. Burstiness $\beta$ equals to zero for a Poisson process (memoryless waiting times, $\rho(\Delta t)=e^{-\Delta t/\mu}/\mu$) and ranges from $-1/2$ (if $\rho(\Delta t)$ is a Dirac distribution) to arbitrarily large positive values (for highly bursty activity). This expression emerges naturally from the dynamical process and can be viewed as a measure of burstiness, equivalent to the commonly used burstiness measure~\cite{Goh08}.

The estimate~(\ref{eq:taumode}) can be further tightened whenever the distribution $\rho(\Delta t)$ of waiting times contains a fat tail, possibly softened by an exponential tail. The archetypical example is a power-law with soft cut-off, $\rho(\Delta t) \propto (\Delta t+A)^{-\gamma} e^{-\Delta t/\tau_\textrm{tail}}$, a frequent model in human dynamics supported by empirical evidence~\cite{Barabasi05, Rocha10, Kivela12}. More generally $\rho(\Delta t) \propto \rho_0(\Delta t) e^{-t/\tau_\textrm{tail}}$, where $\rho_0$ is a fat tailed distribution, that is, decreasing subexponentially. The non-analytic point created in $\rho(s)$ by the fat tail leads to an additional term in the characteristic time (see Methods)
\begin{equation}
\tau_k \approx \max(\mu (|\lambda_k|^{-1} + \beta), \tau_\textrm{tail}).
\label{eq:taumodetail}
\end{equation}
which can be approximated as follows, if $|\lambda_k|^{-1}$ and $\beta$ have different orders of magnitude
\begin{equation}
\tau_k \approx \max(\mu |\lambda_k|^{-1}, \mu \beta, \tau_\textrm{tail}).
\label{eq:taumodemax2}
\end{equation}

\vspace{0.5cm}
\noindent\textbf{Mixing times and dominating mechanisms.} Of particular importance is the slowest non-stationary mode ($k=1$), or mixing mode, the characteristic decay time of which represents the worst-case relaxation time of any initial condition to stationarity. We call this time the mixing time of the process, in generalisation of the classic memoryless case, where it is given by $\tau_\textrm{mix}=\mu \epsilon^{-1}$, determined by the so-called spectral gap \mbox{$\epsilon = -\lambda_1 >0$~\cite{Chung96,Levinmixing}}. This quantity is related to the presence of bottlenecks (that is, weak connections between groups of highly connected nodes, a.k.a.\ network communities) in the network (via Cheeger's inequality~\cite{Cheeger70, Diaconis91}). It is approximated as

\begin{equation}
\tau_\textrm{mix} \approx \max(\mu \epsilon^{-1}, \mu \beta, \tau_\textrm{tail}).
\label{eq:taumodetailmix}
\end{equation}
This expression shows that the asymptotic dynamics of each mode is determined by the competition between three factors: a structural factor ($\mu \epsilon^{-1}$), associated to the spectral properties of the Laplacian of the underlying network, and two temporal factors associated to the shape of the waiting time distribution, namely its burstiness ($\mu \beta$) and its exponential tail ($\tau_\textrm{tail}$). The slowest (largest) of these factors dominates the asymptotic dynamics (Fig.~\ref{fig1}c).

We emphasise that the burstiness and the fat tail effects are not necessarily related~\cite{Lambiotte13}. For instance a power-law $\rho(\Delta t) \propto (\Delta t+1)^{-3}$ restricted to times $\Delta t \leq T$ (sharp cutoff), is arbitrarily bursty but has no tail at all ($\tau_\textrm{tail}=0$) for large but finite $T$. On the other hand, the delayed power-law $\rho(\Delta t) \propto (\Delta t-T)^{-4}$, restricted to times $\Delta t\geq T+1$ has a fat tail (it decreases subexponentially, $\tau_\textrm{tail}=\infty$) but low burstiness $\beta$ for  large $T$, as the mean time $\mu$ increases without bound and the variance remains constant. In the latter case, as for all pure power laws, $\rho(\Delta t) \sim \Delta t^{-\gamma}$ for large $\Delta t$, the mixing time $\tau_\textrm{mix}$, as $\tau_\textrm{tail}$, is actually infinite, reflecting that mixing, or relaxation to stationarity, occurs only with subexponential convergence. In general, the properties of the tail of the distribution depend on the high-order moments of the distribution, and not only on the first two moments as captured in the coefficient of burstiness.

In Fig.~\ref{fig2}, we study the mechanism dominating the mixing time in toy synthetic temporal networks with different waiting time distributions. Not surprisingly, structure is the driving mechanism when waiting times are narrowly distributed around a mean value, as in the case of Dirac (discrete-time, Fig.~\ref{fig2}c) and Erlang distributions (resembling a discrete-time distribution with small fluctuations, Fig.~\ref{fig2}d). For those, the slowdown factor $\Theta=\tau_\textrm{mix}/(\mu \epsilon^{-1})$, comparing the exact mixing time (computed with equation~(\ref{eq:exact}) in Methods), to what it would be with memoryless waiting times ($\mu \epsilon^{-1}$), takes value in $[0.6,1]$. This indicates a limited speed up of the mixing due to negative burstiness, $\beta < 0$, while the structure plays a major role through $\epsilon$ traversing orders of magnitude.

On the other hand, competition between structure and time appears in scenarios of high temporal heterogeneity. For example, only strong communities are able to dominate power-law waiting times (Fig.~\ref{fig2}e-f). The effect of structure on the mixing times is otherwise removed as burstiness (Fig.~\ref{fig2}e) or tail (Fig.~\ref{fig2}f) becomes the leading mechanism, scaling the mixing times up to $14$-fold in the shown configurations. The transitions between the different mechanisms for a range of power-law configurations are presented in Figure~\ref{fig:SI} and Supplementary Note 1.

\vspace{0.5cm}
\noindent\textbf{Model reduction.} The use of coarse-graining through time scale separation, which is the separate treatment of fast and slow dynamics that coexist inside a system,  is crucial to reduce the complexity of systems made of a large number of interacting entities~\cite{simon1961aggregation,koko1999singular,Gfeller}. This procedure is well known for differential equations like~(\ref{eq:diffusion}). In this case, it consists in neglecting fast decaying modes, for example with decay time less than a certain threshold $\tau_\textrm{tresh}$---an approximation invalid for early times but acceptable for times significantly larger than $\tau_\textrm{tresh}$. Only the dominant modes, thus fewer variables, are left in the reduced model. Decreasing the threshold time, one produces a full hierarchy of increasingly more accurate models, but also with increasingly more variables. Reduced models have a clear interpretation from the structural point of view, as fast modes typically correspond to the fast convergence of the probabilities of presence on nodes to a quasi-equilibrium within a network community. This process is followed by a slow equilibration of the population of random walkers trapped in each community to a global equilibrium~\cite{simon1961aggregation,Delvenne13,simonsen2005diffusion}. Each new reduced variable can therefore be interpreted as the  slow-varying probability of presence in each community, as if it had been collapsed into a single node. The hierarchy of increasingly more accurate and complex reduced models corresponds in this structural picture to a hierarchy of increasingly finer partitions into communities. Given that the decay times of the different modes in equation~(\ref{eq:diffusion}) correspond to the Laplacian eigenvalues, the $k$-community partition is unsurprisingly found to be encoded into the $k$ dominant Laplacian eigenvectors, in a way that is decoded by spectral clustering algorithms~\cite{vonluxburg2007,shen2010spectral}.

\begin{figure}[htb]
\centering
\includegraphics[scale=1.]{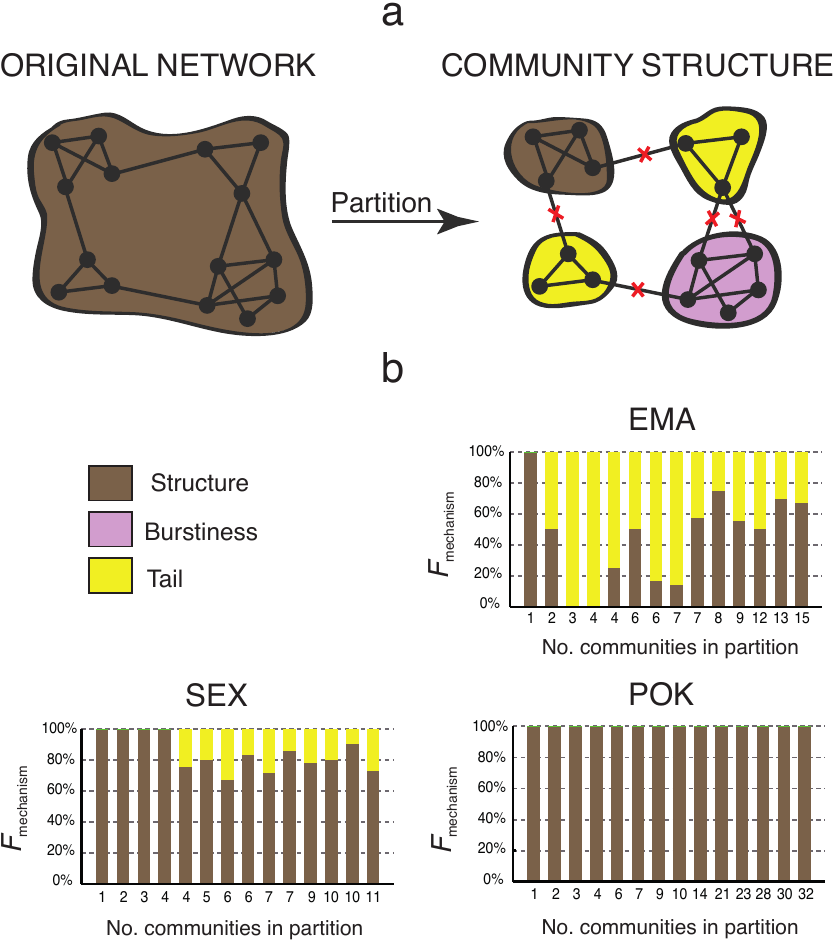}
\caption{\textbf{Global mixing time vs mixing times in communities.} (a) While the mixing time of diffusion on a network may be defined by the community structure, diffusion in the communities themselves may be regulated by structural or temporal patterns. (b) The original network constructed from an empirical dataset is divided into communities with a multi-resolution community detection algorithm, for different values of the resolution parameter (See Methods). For each partition, we identify the fraction of communities larger than $10$ nodes (y-axis) where either structure or tail dominates (according to equation~(\ref{eq:taumodetailmix})), and plot against the number of communities detected at each partition (x-axis). This confirms our spectral analysis, that predicts a compact description of the diffusion dynamics on EMA into three modes describing a probability flow between three communities, each dominated by fat tailed waiting times. Also in agreement with the spectral viewpoint, this is not the case for SEX or POK, unless at the level of a few-node communities.}
\label{fig3}
\end{figure}

This methodology can be extended to more general temporal operators $D$, where the successive decay times now depend on the Laplacian eigenvalues and the temporal characteristics, following equation~(\ref{eq:taumodemax2}). However, the effect of temporality may have non-trivial consequences, as it may limit the number of reduced models. As an extreme scenario, let us consider a complex network in which the transient dynamics is dominated by the fat tail of the waiting time distribution, such that all transient modes decay for large times as $e^{-t/\tau_\textrm{tail}}$. Following equation~(\ref{eq:eigenmodes}), the approach to stationarity is described by the superposition of modes, all of the form $f_k(t) e^{-t/\tau_\textrm{tail}} \bm{\mathrm{v}}_k$, for various functions $f_k(t)$ that decay more slowly than exponentially, as for example $f_k(t)\propto t^{-\gamma_k}$. Among those, any mode can only become negligible with respect to another after a very long time. For all practical purposes, the system exhibits an inherently complex dynamics with no time scale separation, as the fine scale structure (the finest being at the level of a single node) and the large scale network communities have equal or similar impact on the random walk dynamics. As a consequence, only two reduced models are available, one in which all $N$ modes are considered, and another described solely by the stationary state. 

In general, only the Laplacian eigenvalues smaller in magnitude than $\mu/\tau_\textrm{tail}$ determine corresponding decay times. Just like in classical memoryless case, they generate a hierarchy of times scales and reduced models corresponding to multiple levels of community structures. For instance, if 5 eigenvalues are smaller than $\mu/\tau_\textrm{tail}$, thus $0=\lambda_0 > \lambda_1 > \cdots > \lambda_4 >-\mu/\tau_\textrm{tail}$, then there is a reduced diffusion model capturing the movement of the random walker between 5 aggregated nodes. The mixing, internal to each community and decaying as $e^{-t/\tau_\textrm{tail}}$, is considered instantaneous at large enough time scales. Coarser aggregation, for example based on two communities and two eigenvalues $\lambda_0, \lambda_1$, may be relevant although valid only at even larger times. However, aggregation based on any finer partitioning (other than into one-node communities) has the same accuracy as the five-community model, and thus little practical value as a reduced model. The degeneracy of characteristic decay times therefore limits the number of useful reduced models. This implies that a decomposition into communities is not necessarily associated to a time scale separation, or a reduced model, of the dynamics. For this reason, only models incorporating 1 to 5 or $N$ dominating modes are adequate for this particular example. This smaller choice of reduced models has ambivalent consequences. It limits the set of resolutions at which to describe the dynamics, but also provides a natural level for community structure (an open problem in multi-scale community detection~\cite{Fortunato10,ronhovde,Delmotte2011,karrer,reichardt}), defined as the finest partitioning yielding a reduced model for the system. 

\vspace{0.5cm}
\noindent\textbf{Numerical Analysis.} Equation~(\ref{eq:taumodetailmix}) shows that the mixing time depends in first approximation only on the mean, variance and tail of the waiting time distribution, while other properties of the distribution are irrelevant. With numerical simulations, we validate this approximation and provide some quantitative intuition on the competition between the three factors regulating the mixing. We construct synthetic temporal networks respecting our assumptions, for example stationarity or absence of correlations, and calibrate them with the static structure and the inter-contact time distribution observed on a number of empirical datasets.  The empirical networks used for calibration correspond to face-to-face interactions between visitors in a museum (SPM), between conference attendees (SPC)~\cite{Isella11}, and between hospital staff (SPH)~\cite{Vanhems13}; email communication within a university (EMA)~\cite{Eckmann04}; sexual contacts between sex-workers and their clients (SEX)~\cite{Rocha10}; and communication between members of a dating site (POK)~\cite{Holme04} (See Methods). On these networks we observe the waiting time distribution and the spectral gap, which allows to compute both the exact mixing time, with equation~(\ref{eq:exact}) in Methods, and its approximation in equation~(\ref{eq:taumodetailmix}). The results, reported in Table~\ref{tab01}, show a good agreement, except for SPC, as analysed later in this section.

\begin{table}[h]
    \begin{tabular}{|c|cccccc|}
    \hline
    ~                                           & SPM  & SPC  & SPH  & EMA  & SEX  & POK   \\ \hline
    $\Theta$                               & 1.77 & 1.81 & 1.54 & 1.19 & 1.02 & 1.02  \\
    Exact Mixing                         & 54   & 205  & 671  & 718  & 7,887 & 29,347 \\
    Approximate Mixing              & 61   & 313  & 664  & 603  & 7,741 & 28,685 \\
    Relative Difference (\%)        & 13.0 & 52.7 & -1.0 & -16  & -1.9 & -2.3  \\ \hline
    \end{tabular}
\caption{\textbf{Exact vs. approximate mixing times.} The table shows that for six configurations of temporal networks the slowdown factor $\Theta=\tau_{\textrm{mix}}/(\mu \epsilon^{-1})$, that is the ratio of the exact mixing time, calculated from equation (\ref{eq:exact}) using the simulated waiting time distribution, to the mixing time we would have for the exponential case (Poisson process, with same mean). The results indicate strong (SPM, SPC and SPH), medium (EMA) and weak (SEX and POK) slowdown ($\Theta>1$). The approximate mixing time, computed with equation (\ref{eq:taumodetailmix}), shows a good agreement with the exact value. }  
\label{tab01}
\end{table}

Except for SEX~\cite{Rocha11} and POK datasets, in the other cases the temporal heterogeneity substantially increases the mixing times (see slowdown factor $\Theta$ in Table~\ref{tab01}). Fig.~\ref{fig3arxiv} shows that in these networks the dominant factor regulating the mixing time depends on the characteristics of the system. Fat tails of the waiting time distribution drive the relaxation for the cases of face-to-face contacts (SPM, SPC, SPH). Structure however is the leading mechanism behind the networks corresponding to other situations of human communication (EMA, SEX, POK). 

If communities are completely removed by randomising the network structure, the link sparsity of SEX and POK networks guarantees that structure remains dominating, as the sparsity results in the inevitable creation of bottlenecks for diffusion even in a random network. On the other hand, in the EMA network, which has a relatively dense connected structure, absence of communities leads to temporal dominance (Fig.~\ref{fig3arxiv}). Finally, when the contact times are uniformly distributed, we recover the well-known result that network structure, with or without communities (Fig.~\ref{fig3arxiv}), is the main factor regulating the convergence to stationarity.

Because the raw empirical temporal networks do not necessarily abide by our simplifying assumptions, one cannot  validate our theoretical derivations, for example, by estimating the mixing time directly from simulations on empirical data. Indeed the periodic rythms  or correlations between successive jumps or other temporal patterns may induce effects not captured by our formula, as discussed in  Table~\ref{tab_1_SI} and Supplementary Note 3. 

Empirical data are collected during a finite time span $T$, leaving the choice between a model for $\rho(\Delta t)$ that is limited to $T$ (sharp cut-off) or extrapolated to infinite times with a tail decay (soft cut-off). The latter may occur for instance if we have theoretical or practical reasons to prefer a fat-tail-based model (for instance power-law) with exponential cut-off for $\rho(\Delta t)$. This choice is of little impact on the results, provided a sufficient observation time $T$ (Supplementary Note 2). When temporal patterns trump structure, we have a self consistency condition $\tau_\textrm{mix} \approx \mu(\epsilon^{-1}+\beta) \approx \tau_\textrm{tail}$  (Fig.~\ref{fig3arxiv}) for SPM and SPH), expressing that both models lead to similar mixing times, albeit through the different mechanisms of either burstiness (sharp) or fat-tail (soft). For SPC, self consistency is not attained. This happens possibly because the observation time of $\sim$ 2 days is not sufficiently large to dilute the irregularity on the observed inter-contact time distribution (thus in the simulated waiting times) induced by the inactivity during night periods. Consequently, this makes the tail difficult to measure as it has not clearly emerged from a still transient behaviour, and a significant mismatch is observed between exact and approximate mixing (Fig.~\ref{fig3arxiv}).

We evaluate the possible sizes of reduced models for the data as follows. As the eigenvalues $\lambda_k$ of the Laplacian range between $0$ and $-2$, the number of modes with structurally determined decay times  ($-\mu/\tau_\textrm{tail} < \lambda_k \leq 0$) can be roughly evaluated to $N \mu/2 \tau_\textrm{tail}$  on an $N$-node network, if eigenvalues are evenly spaced. A similar reasoning holds whenever burstiness dominates the tail effect.  This analysis reveals three different scenarios in the structure-dominated  datasets considered above: i) SEX benefits from a full hierarchy of reduced models, as $\mu/2\tau_\textrm{tail} >1$. ii) The dynamics on EMA can be approximated with just three modes, associated to three network communities, as $0=\lambda_0 > \lambda_1 >\lambda_2 > -\mu/\tau_\textrm{tail} > \lambda_3 > \ldots$, while a more detailed description, unless at the node level, would not gain any significant accuracy as further modes are all degenerate with the same decay time $\mu/\tau_\textrm{tail}$. iii) POK exhibits a practically full hierarchy of reduced models as around one third of its modes are determined by structure, $\mu/2\tau_\textrm{tail}=0.33$, therefore the finest reduced model is at a  few-node community level. This spectral analysis is comforted by applying a multiscale community detection algorithm to the empirical networks, which finds a partition into three communities dominated by fat-tail effects for EMA, but not for POK or SEX (Fig.~\ref{fig3}). 

\section*{\large Discussion}

We have presented a unified mathematical framework to calculate the relaxation time to equilibrium in a wide variety of stochastic processes on networked temporal systems. Our formalism is able to refer to arbitrary linear multi-agent complex systems, including linearisations of non-linear dynamical models such as Kuramoto oscillators or  non-Laplacian   diffusion dynamics such as SIR-like epidemics, as detailed in Methods. It is also possible to accommodate non-uniformity of parameters as nodes are not identical in real systems. Our results are particularly relevant to improve the understanding of temporal networks, by highlighting the important interplay between structure and the temporal statistics of the network. Our formalism is different from previous studies of random walks on temporal networks that focus on homogeneous temporal patterns~\cite{Perra2012} or do not account for the competition between structure and time~\cite{Starnini12}. We emphasise that questions related to ordering, not timing of events, such as the number of hops before stationarity or the succession of nodes most probably traversed by the walker, depend on the structure alone and not of hops timing statistics. Moreover, key mechanisms have been left aside in our modelling approach, such as the non-stationarity or periodicity of most empirical networks~\cite{Rocha13, Holme13, Horvath14}, and the existence of correlations between edge activations and therefore preferred pathways of diffusion~\cite{Kivela12, Scholtes2014, Rosvall2014}. Important future work includes their incorporation in our mathematical framework, and the identification of dominant mechanisms in empirical data.

We have shown that, in the absence of some temporal correlations, the characteristic times of the dynamics are dominated either by temporal or by structural heterogeneities, as those observed in real-life systems. The competing factors are not only observed in different classes of networks modelled from empirical data but also at different hierarchical levels of the same network represented by its different scales of community structure. We have also identified two contrasting metrics of the statistics of waiting times, burstiness and fat-tails. We have shown that they regulate the dynamics on the network in different ways. In systems where temporal patterns are the dominating factor, the reduced models obtained by aggregation of communities as commonly used in practice  are not necessarily relevant, as small scale details are impenetrably intertwined with large scale structure to form a complex global dynamics. In general, the temporal characteristics impose the natural description levels of the dynamics.  Altogether, our results suggest the need for a critical assessment of a complexity/accuracy trade-off when modelling network dynamics. In some classes of real-world systems, the burden of increased model complexity may not compensate the incremental gain in knowledge, while other systems require the fine network structure as a key ingredient to any realistic modelling.

\begin{footnotesize}
\section*{\large Methods}
\noindent\textbf{Overview.} 
We provide a description of multi-agent linear dynamics, detail the derivation of approximation~(\ref{eq:taumodetail}) and its range of validity, illustrate the generality of the framework beyond random walks, and describe the empirical data and numerical implementations. In the Supplementary Information, we discuss on power-laws and cut-offs (Supplementary Note 1 and Figure~\ref{fig:SI}),  the evaluation of waiting time distributions in empirical data (Supplementary Note 2), and  the adequacy of our theoretical framework in empirical networks with correlations and non-stationarity (Supplementary Note 3). \\

\noindent\textbf{Linear multi-agent systems.}  
%\label{SI_sec1}
We derive equation~(\ref{eq:DxLx}) for general linear networked systems, or multi-agent systems~\cite{Fax04}, with an illustration on consensus dynamics. Every node $i$ is modelled by a linear agent whose internal state is initially zero and that  converts an input signal $u_i$ with an operator $F_i$---the so-called transfer function---into a state signal $x_i=F_i u_i$. Here, $u_i$ represents a time trajectory $u_i(t)$ or its Laplace transform $u_i(s)$, and similarly for the state trajectory $x_i$. The transfer function $F_i$ is a operator mapping input trajectories $u_i$ to state trajectories $x_i$, that is requested to be linear, causal (if $u_i(t)=0$ for all $t <T$, then the same holds for $x_i(t)$), and time invariant (shifting $u_i(t)$ in time results in shifting $x_i(t)$). Under those conditions, this operator takes the form $x_i(s)=F_i(s) u_i(s)$, a simple product of functions, in the Laplace domain.

When using Laplace transforms it is customary  to use time domain trajectories that are zero for negative times. We account for the state initial condition $x_i(0)=x_{i,0}$ with another input term $v_i$ that can be for example an impulse exciting the rest state to any desired initial condition at time zero. The agent dynamics therefore writes $x_i(s)=F_i(s)(u_i(s) + v_i(s))$. For instance if $F_i$ is the integration operator in the time domain, $x_i(t) = x_{i,0}+\int_0^t u_i(t') \dif t'$ from the initial value at time $0$, or in other terms $x_i(t) = \int_0^t u_i(t') + x_{i,0}\delta(t) \dif t'$ (for the Dirac impulse $\delta(t)$) then in the Laplace domain we find $x_i(s)= 1/s(u_i(s) + x_{i,0})$; in this case the Laplace domain operator is the multiplication by $1/s$, and $v_i$ is a Dirac impulse in the time domain, a constant in the Laplace domain that encodes the initial condition. One can as well invert the transfer operator, $D_i=F_i^{-1}$,  and write $\dif x_i/\dif t=u_i(t)+x_{i,0}\delta(t) $, or $s x_i(s)= u_i(s) + x_{i,0}$; here the Dirac impulse is seen as arising from differentiating the discontinuous step of the state from rest to initial condition $x_{i,0}$.

Connecting agents such that the input $u_i(t)$ of agent $i$ is a weighted sum of other agents' states $\sum_j L_{ij} x_j(t)$ leads to a global dynamics  $\dif \bm{\mathrm{x}}/\dif t=L\bm{\mathrm{x}}+\bm{\mathrm{x}_{0}} \delta(t)$, or equivalently $s\bm{\mathrm{x}}(s)=L\bm{\mathrm{x}}(s)+\bm{\mathrm{x}_{0}}$, where $L$ is the matrix of entries $L_{ij}$, $\bm{\mathrm{x}}$ is the vector of states $x_i$, and $\bm{\mathrm{x}_{0}}$ the vector of initial conditions $x_{i,0}$. When $L$ is a Laplacian (rows summing to zero, nonnegative off-diagonal weights), this is a simple consensus system where agents (for instance robots or individuals) change their state (for instance position, opinion modelled as a real number) towards an average of their neighbours' states. With arbitrary interconnection weights and arbitrary transfer functions, one similarly obtains

\begin{equation}
D\bm{\mathrm{x}}=L\bm{\mathrm{x}}+\bm{\mathrm{v}}
\label{eq:M1bis}
\end{equation}
where $D$ is the diagonal matrix of transfer functions $F_i^{-1}$ ($D$ reduces to a scalar in case of identical agents $F_i=F_j$), and $\bm{\mathrm{v}}$ the vector of initial condition input $v_i$. The case when $L$ remains Laplacian, but $D$ is arbitrary, describes general, higher-order, consensus systems where the agents converge to equal values on a strongly connected network through arbitrarily complicated internal dynamics, for example modelling realistic robots or vehicles. For example, vehicles of mass $m$ driven by a force and friction will obey $m \bm{\ddot \mathrm{x}} + \alpha \bm{\dot \mathrm{x}}=L\bm{\mathrm{x}}$. An abuse of notation allows to drop implicitly the initial condition and write simply $\dif \bm{\mathrm{x}}/\dif t=L\bm{\mathrm{x}}$, and more generally equation~(\ref{eq:DxLx}), $D\bm{\mathrm{x}}=L\bm{\mathrm{x}}$, instead of equation~(\ref{eq:M1bis}) above. 
Discrete-time systems are recovered with a discrete-derivative operator $D\bm{\mathrm{x}}(t)=\bm{\mathrm{x}}(t+1)-\bm{\mathrm{x}}(t)$. 

\noindent\textbf{Exact and approximate characteristic times for random walkers.}
We derive the approximate characteristic times (\ref{eq:taumodetail}) of decaying modes associated to a general random walk. The dominant mode associated to $\lambda_0=0$ is the stationary distribution, unique for an ergodic random walk. The relaxation time to the stationary distribution, generalising the well-known mixing time of Markov chain theory~\cite{Levinmixing}, is the characteristic time of the slowest decaying mode after the stationary mode, usually associated to $\lambda_1$, as we suppose here. An example of a case when $\lambda_1$ is not associated to the mixing mode, but rather $\lambda_N$, is the discrete-random walk, where the distribution is a Dirac distribution at $\mu$, with zero variance, and the network is bipartite or close to bipartite. 

The characteristic decay time of a time-domain function $f(t) \sim e^{-t/\tau_\textrm{decay}}$ can be found in the Laplace domain, as $f(s)$ is defined and analytic over all $s$ of real part larger than $-1/\tau_\textrm{decay}$, but undefined or non-analytic in at least one point $s$ of real part $-1/\tau_\textrm{decay}$. For example the Laplace transform of $e^{-t/\tau_\textrm{decay}}$ is $1/(1+\tau_\textrm{decay}s)$, with pole at $s=-1/\tau_\textrm{decay}$, and the Laplace transform of $(t+1)^{-\gamma}e^{-t/\tau_\textrm{decay}}$ for $\gamma > 1$, is defined, but non-analytic, at \mbox{$s= -1/\tau_\textrm{decay}$.}  

The eigenfunction $z_k(t)$ associated to $\lambda_k$ is the solution of $D(s) z_k(s)=\lambda_k z_k(s) + D(s) z_k(t=0)/s$, derived in the text as equation~(\ref{eq:M3}), with $D(s)=1/\rho(s)-1$ and $L$ replaced with $\lambda_k$. The decay time of $z_k(t)$ is found if we find 
the right-most non-analytic point of 

\begin{equation}
z_k(s)=\frac{1}{D(s)-\lambda_k}\frac{D(s)}{s} z_k(t=0),
\label{eq:2222}
\end{equation}
derived in the main text as equation~(\ref{eq:rwmodes}). Note that for $\lambda_k \neq 0$, the singularity in $s=0$ is only apparent as $D(s)=\mu s - (\sigma^2-\mu^2) s^2/2 + \ldots$  This expression is non-analytic in $s$ in exactly two circumstances: i. $D(s)$ is non-meromorphic in $s$, meaning that it cannot be defined analytically in a punctured neighbourhood of $s$ (a neighbourhood of $s$, minus $s$) or ii. $D(s)=\lambda_k$.
 
Case i. happens when $\rho(s)$ is non-meromorphic in $s$. A typical example is when $\rho(\Delta t)\propto \rho_0(\Delta t)e^{-\Delta t/\tau_\textrm{tail}}$, where $\rho_0(t)$ is a fat-tailed distribution, in other words decreases more slowly than any exponential, leading to a non-meromorphic transfer function---a rare occurrence in standard systems theory. The non-analytic point is precisely $s=-1/\tau_\textrm{tail}$. 

Case ii. commands to solve the equation 
\begin{equation}
\rho(s)=1/(1+\lambda_k), 
\end{equation}
whose solution $s_k$ enters the expression for the exact characteristic decay time, combining the two cases:
\begin{equation}
\tau_{k}=\max(-1/s_k,\tau_\textrm{tail}),
\label{eq:exact}
\end{equation}
which can be approximated from an expansion of $\rho(s)$. Note that $\rho(s)=1-\mu s + (\mu^2+\sigma^2) s^2/2 - \cdots$ is also called the moment generating function, whose successive derivatives around $s=0$ are the moments of the distribution, up to sign. 
We consider the Pad\'e (that is, rational) approximation:
\begin{equation}
\rho(s) \approx \frac{2\mu + (\sigma^2-\mu^2)s}{2\mu + (\sigma^2+\mu^2)s},
\label{eq:paderho}
\end{equation}
equivalent for small $s$ to a second-order Taylor approximation in terms of accuracy. This approximates the transfer function $D^{-1}(s)=F(s)= \rho(s)/(1-\rho(s))$ as
\begin{equation}
F(s) \approx \frac{1}{\mu s} + \beta,
\label{eq:padeF}
\end{equation}
where the adimensional term $\beta=\frac{\sigma^2-\mu^2}{2\mu^2}$ takes its minimum value at $-1/2$ for the discrete-time random walk, zero for a memoryless process, and arbitrary large values for highly heterogeneous distributions.  It is therefore a suitable measure of the burstiness of the process~\cite{Goh08,Kivela12}. Such a transfer function is called Proportional-Integral, a common class in systems theory~\cite{astrom2010feedback}. The equation $D(s)=\lambda_k$ is approximately solved by $1/s_k \approx \mu(|\lambda_k|^{-1}+\beta)$. A possible non-analytic point at $s=-1/\tau_\textrm{tail}$ caps the characteristic decay time of eigenfunction $k$ to

\begin{equation}
\tau_k \approx \max(\mu (|\lambda_k|^{-1} + \beta), \tau_\textrm{tail}).
\tag{\ref{eq:taumodetail}}%\label{eq:apprchartime}
\end{equation}

\begin{figure}[htb]
\centering
\includegraphics[width=1.\columnwidth]{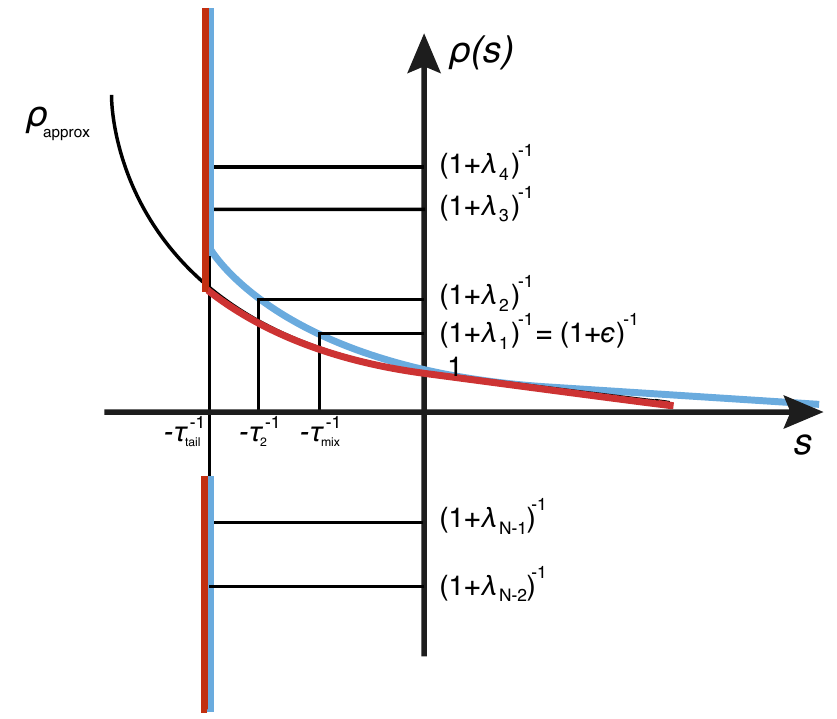}
\caption{\textbf{The exact and approximate characteristic times of a waiting time probability distribution.} Computation of the characteristic time of the distribution $\rho(s) \propto \rho_0(t) e^{-t/\tau_{\textrm{tail}}}$ (blue curve) with a fat-tailed $\rho_0$ and a non-analytic point at $s=-1/\tau_{\textrm{tail}}$. The exact characteristic decay times $\tau_k$ are found from the eigenvalues and the blue curve, following equation~(\ref{eq:exact}). In this example only three modes, including the stationary, are influenced by the structure of the network, while the other modes collapse to a single decay time $\tau_{\textrm{tail}}$. The dynamics can thus be approximated by three modes, typically describing aggregate probability flows between three network communities. The approximate solution equation~(\ref{eq:taumodetail}) follows the red curve, constructed from the Pad\'e approximant $\rho_\textrm{approx}$, see equation (\ref{eq:paderho}). As eigenvalues $\lambda_k$ change from $0$ to $-2$, the values $1/(1+\lambda_k)$ can take any real value larger than $1$ or less than $-1$.}
\label{fig4}
\end{figure}

If $\lambda_k^{-1}$ and  $\beta$ are of different orders of magnitude, one may further approximate as

\begin{equation}
\tau_k \approx \max(\mu |\lambda_k|^{-1}, \mu \beta, \tau_\textrm{tail}).
\tag{\ref{eq:taumodemax2}}%\label{eq:apprchartime2}
\end{equation}
in which the influence of the fat tail, the structure and the burstiness are decoupled. The mixing time is associated with $\lambda_1$. Fig.~\ref{fig4} shows that typically half of the modes have structurally determined decay times, namely those with positive $1+\lambda_k$, if eigenvalues are uniformly spaced between $0$ and $-2$, in apparent contradiction with the back of the envelope calculation $N\mu/2\tau_\textrm{tail}$ proposed in the main text, which can reach $N$. This is a consequence of the progressive degradation of the Pad\'e approximation for larger $s$ (larger eigenvalues).

Even for small eigenvalues, for example for $\lambda_1=-\epsilon$, the approximation (\ref{eq:taumodetail}), thus also  equation~(\ref{eq:taumodetailmix}), has a limited range of validity. The underlying Pad\'e approximation (\ref{eq:paderho}) is valid for small enough $s$, while for large enough $s$ the exact solution is $\tau_\textrm{tail}$. The small $s$ behaviour is captured by the first and second moments, and the large $s$ behaviour, in other terms the tail $\tau_\textrm{tail}$, determines the growth of high order moments. This double approximation based on high and low moments may explain why the approximation is successful on diverse data sets (see Fig.~\ref{fig3}a). However it is likely to fail precisely when the intermediate moments  (third, fourth, etc.), not covered by the approximation, dominate the behaviour of the moment generating function $\rho(s)$. This occurs for instance in the case of a power law of exponent $>3$ with sharp cut-off at a large time: while the first and second moment remain bounded and the tail is absent, the intermediate moments grow without bound with the cut-off time. Such distributions are rarely used as models for real-life data however. On the other hand, numerical experiments showed excellent accuracy for a wide range of power-laws with soft cut-off.

\noindent\textbf{Beyond random walks.}
Random walks  are many times formally identical to the asymptotic behaviour of non-linear dynamical models, and also serve as a prototype for a wider class of dynamics that includes epidemic spreading. For example, power networks may be modelled as a network of second-order Kuramoto oscillators~\cite{dorfler2012synchronization} whose state in every node is a phase (an angle) $\theta_i$ influenced by its neighbours 
\begin{equation}
I \ddot \theta_i + \alpha \dot \theta_i = \omega + \sum_{j} L_{ij} \sin(\theta_j-\theta_i),
\label{eq:Kura}
\end{equation}
where $I$ is the inertia, $\alpha$ the friction, and $\omega$ describes the natural frequency of the oscillation. In this construction, we assume identical parameters for each node. After the change of variables $\phi_i=\theta_i - \omega t/\alpha$, the linearisation around the synchronised equilibrium ($\phi_i=\phi_j$ is constant for all times) is written
\begin{equation}
I  \Delta \ddot \phi_i + \alpha \Delta \dot \phi_i =  \sum_{j} L_{ij} \Delta\phi_j,
\label{eq:Kura2}
\end{equation}
which is formally identical to a consensus equation structured by the Laplacian $L$~\cite{dorfler2012synchronization}. Hence it supports the same formalism as random walks, after the operator $D$ is changed accordingly. The linear dynamics associated to the asymptotic behaviour is often a crucial first step in understanding the non-linear dynamics of the network at some time scale.

Another class of dynamical systems where our results apply are linear dynamics given by equation~(\ref{eq:DxLx}), but where the interaction matrix $L$ does not have a Laplacian structure, and where the dominant eigenmode is not necessarily stationary. This is the case, for instance, for epidemic processes where infected individuals diffuse on a meta-population network, where nodes are large populations (cities, etc.), and the individuals may either  reproduce (by contamination of a healthy individual) or die (or recover). One classic model for this process is a multi-type branching process.  This is akin to a random walk whose number of walkers is not preserved in time, as random events are not limited to hops but also death or split. This kind of model also emerges as the linearisation of classic compartmental epidemic models such as SI or SIR~\cite{keeling2008methods}. The waiting time between two consecutive events may also be described by an arbitrary distribution $\rho(\Delta  t)$. The expected number of walkers in time may again be described by an equation $D\bm{\mathrm{x}}=L\bm{\mathrm{x}}$, where $D$ depends on $\rho$ in the same way as before, but now $L$ is no more a Laplacian, and in particular may have negative and positive eigenvalues. If the epidemics is supercritical, then the system is unstable. This implies the existence of unstable eigenmodes ($\lambda_k > 0$) where the number of walkers grows exponentially as $e^{t/\tau_{k}}$, with $\tau_{k}$ now approximated by $\mu(-\lambda_k^{-1}-\beta)$. In this case, the fat tail of the waiting time distribution does not play any role for the unstable eigenmodes. From this formula, we see that while the decay of stable eigenmodes tends to slow down due to burstiness and possibly fat tail, the unstable eigenmodes are boosted by burstiness. As a consequence, it leads in all eigenmodes to a long-term increase of the infected population due to the temporal heterogeneity.

\noindent\textbf{Numerical analysis.} We use six datasets of empirical networks: face-to-face interactions between visitors in a museum (SPM), between conference attendees (SPC)~\cite{Isella11}, and between hospital staff (SPH)~\cite{Vanhems13}; email communication within a university (EMA)~\cite{Eckmann04}; sexual contacts between sex-workers and their clients (SEX)~\cite{Rocha10}; and online communication between members of a dating site (POK)~\cite{Holme04} (See Table~\ref{tab03}).

\begin{table}[h]
\centering
\begin{tabular}{|c|cccccc|}
\hline
        & SPM & SPC & SPH & EMA & SEX & POK \\
\hline
$N$ & 72 & 113 & 75 & 3,186 & 11,416 & 28,295 \\
$L$ & 6,980 & 20,818 & 32,424 & 309,120 & 33,645 & 528,869 \\
$\delta t$ & min & min & min & hour & day & hour \\
$T$ (days) & 1 & 2.5 & 4 & 82 & 1,000 & 512 \\
$\mu$    &  15.4 & 80.5 & 288 & 61.5   &   92.9  & 1,205  \\
$\sigma$  & 30.8  &  143 & 502 &   119   &  121 &  1,634  \\ 
$\epsilon$ & 0.505 & 0.709 & 0.66 & 0.102 & 0.012 & 0.042 \\
$\tau_\textrm{tail}$ & 61 & 313   & 664  & 207   & 98    & 1,755  \\
\hline
\end{tabular}
\caption{\textbf{Summary statistics of the empirical networks.} For the largest connected component of the empirical networks used in this study we display the number of vertices ($N$); number of links ($L$); temporal resolution used in the simulations ($\delta t$);  total duration of the network data ($T$); average ($\mu$) and standard deviation ($\sigma$) of waiting times for the simulated random walk, in same unit as $\delta t$; and the spectral gap of the original unweighted network ($\epsilon$).}  
\label{tab03}
\end{table}

The synthetic networks used in Fig.~\ref{fig2} are formed by $1000$ nodes and $10000$ links equally divided into two groups, initially disconnected (rewiring probability equal to zero). Within each group, pairs of nodes are formed between uniformly chosen nodes. A fraction of links are rewired to weaken communities. Rewiring consists of uniformly choosing two pairs of nodes and swapping the pairs. Rewiring probability equal to one removes any significant community structure of the network. Communities refer to groups of nodes with more connections between themselves than with nodes at other groups. The spectral gap is calculated for each network configuration. The exemplified waiting time distribution are $\rho(\Delta t)= \exp(-\Delta t/\mu)/\mu$ with $\mu=1$ (exponential); $=1.0$ (discrete); $\propto \Delta t^{k-1}\exp(-k\Delta t/\mu)$ with $k=3$ and $\mu=3$ (Erlang law); $\propto \Delta t^{\gamma}$ with $\gamma=2$ and cutoff at $T=1000$ (power-law with sharp cut-off); $\propto \Delta t^{\gamma}\exp(-\Delta t/\tau_\textrm{tail})$ with $\gamma = 3$ and $\tau_\textrm{tail}=20$ (power-law with soft cut-off).

The synthetic temporal networks used in Figs~\ref{fig3arxiv} and \ref{fig3} are constructed by using the unweighted version of the empirical networks as underlying fixed structures. The randomised networks used in Fig.~\ref{fig3arxiv} are obtained by uniformly selecting two pairs of nodes of the original network and then swapping the respective contacts. In the dynamic network, a node and its links alternate between inactive and active states. In order to fit this synthetic temporal network to our theoretical assumptions, we assume that all links of a node are activated simultaneously, the system is time-invariant (no daily or weekly cycles), exhibits no temporal correlations between successive jumps and node activation times are sampled from the same distribution of inter-activation times (node homogeneity). The active state of a node and its links lasts for one time step $\delta t$ and then the node and links return to the inactive state. The distribution of inter-activation times (a.k.a.\ inter-contact times) corresponding to the \textit{original times} is obtained by pooling the times between two subsequent activations of the same node, in other words two consecutive contacts established between the node and its neighbourhood, as observed in the empirical networks~\cite{Rocha11,Rocha13}. The \textit{randomised times} in Fig.~\ref{fig3arxiv} correspond to activation times sampled from exponential distributions with the same mean as the corresponding empirical cases. To obtain the waiting time estimates presented in Fig.~\ref{fig3arxiv}, we simulate a random walk in these synthetic networks. A random walker starts in a randomly chosen node. As time goes by, a walker remains in the node until its new activation and then hops to a uniformly chosen neighbour. The waiting time is thus the time elapsed between the arrival and departure of the walker in a node. We simulate a single random walker and let it hop 300,000 times, starting at 10 uniformly chosen nodes, hence we collect a total of 3 million points for the statistics of waiting times. Besides that, for each starting configuration, we discard the initial 5,000 hops to remove the stochastic transient. The waiting time distribution observed in the simulations provides us with estimates of $\tau_\textrm{mix}$, $\mu$, $\sigma^2$ and $\tau_\textrm{tail}$. The mixing time $\tau_\textrm{mix}$ is estimated from the distribution expressed in the Laplace domain through equation~(\ref{eq:exact}) (for $k=1$, the mixing mode). This equation is exact since the synthetic data has been constructed so as to satisfy the conditions of validity under which it is proved. One could also estimate the mixing time from the convergence of random trajectories starting in a definite node towards stationarity, however this exercise is computationally demanding. The exponential tail decay time $\tau_\textrm{tail}$ is estimated using least-square fitting by a line in a lin-log plot of the second half of the waiting time distribution data, thus between $T_\textrm{max}/2$ and $T_\textrm{max}$, where $T_\textrm{max}$ is the largest observed waiting time. Indeed an exponential decay translates into a straight line in a lin-log plot, the slope of which determines the decay time. The choice of the interval $[T_\textrm{max}/2, T_\textrm{max}]$ to perform a linear fit is steered by the want for a simple criterion, uniform across datasets. This may be inappropriate for datasets where the observation time forces a cut-off on the datasets before the tail decay has had time to emerge, for instance in SPC, where a more careful definition of tail, for example on a smaller interval, may be more adequate.

Let us interpret some elementary observations that can be made on Fig.~\ref{fig3arxiv}. From a) to b), $\tau_\textrm{tail}$ is only approximately preserved because it is recomputed from the simulated waiting time distribution on a different network, thus a slightly different distribution, even though the inter-contact times are sampled from the same distribution in both panels. In c) and d) the inter-contact time distribution is memoryless (exponential), thus identical to the waiting time distribution regardless of the network, and $\tau_\textrm{tail}$ coincides with $\mu$. From a) to c), the quantity $\mu \epsilon^{-1}$ drops sharply, although $\epsilon$ is clearly preserved as the network is unaltered. The reason is that although the mean inter-contact time is preserved by construction, the mean waiting time also depends on the variance of the inter-contact times, in virtue of the bus paradox, and this variance is clearly strongly affected by the time randomisation. The drop in $\mu \epsilon^{-1}$ from  b) to d) is associated to the same mechanism.

To identify the community structure of the empirical networks in Fig.~\ref{fig3}, we use the partition stability  method~\cite{Delvenne13} using a freely available implementation~\cite{StabilityToolbox}. The method uses a simple diffusion process such that different communities are detected at different time scales according to the potential of the communities to trap the diffusion at the given time scale. Optimal communities have been derived for values of the resolution parameter decreasing from $10^{2}, 10^{1.95}, 10^{1.9},10^{1.85}, \ldots$, with increasingly fine partitions. After identifying the relevant communities, we discard the inter-community links  and calculate the spectral gap of each individual community  of at least 10 nodes. 

\section*{\large Acknowledgements} The authors thank Michael Schaub for carefully reading the manuscript. JCD and RL thank financial support of IAP DYSCO and ARC `Mining and Optimization of Big Data Models'. RL acknowledges support from FNRS and from FP7 project `Optimizr'. LECR is a FNRS charg\'{e} de recherches and thanks VR for financial support. 

\section*{\large Author Contributions} JCD, LECR, and RL conceived the project and wrote the manuscript. JCD derived the analytical results. LECR performed the numerical simulations and analysed the data. 

\section*{\large Additional information} The authors declare that they have no competing financial interests. Correspondence and requests for materials should be addressed to JCD~\mbox{(Jean-Charles.Delvenne@uclouvain.be)}.

\newpage
\appendix

\newpage
\section*{SUPPLEMENTARY NOTE 1}
\label{I}

We extend the discussion on the main text on power-law distribution of waiting times and show results for various values of parameters, including the transition thresholds between dominating factors. Power-law functions represent a particularly important family of probability distributions. Although many times power-laws are not the optimal statistical probability distribution to model waiting times of human interactions, they properly capture both aspects of realistic distributions, i.e.\ fat tails and burstiness, in a single functional form, and is in contrast to standard models of homogeneous (e.g.\ exponential) and regular times. 

\begin{figure}[H]
\centering
\includegraphics[scale=0.9]{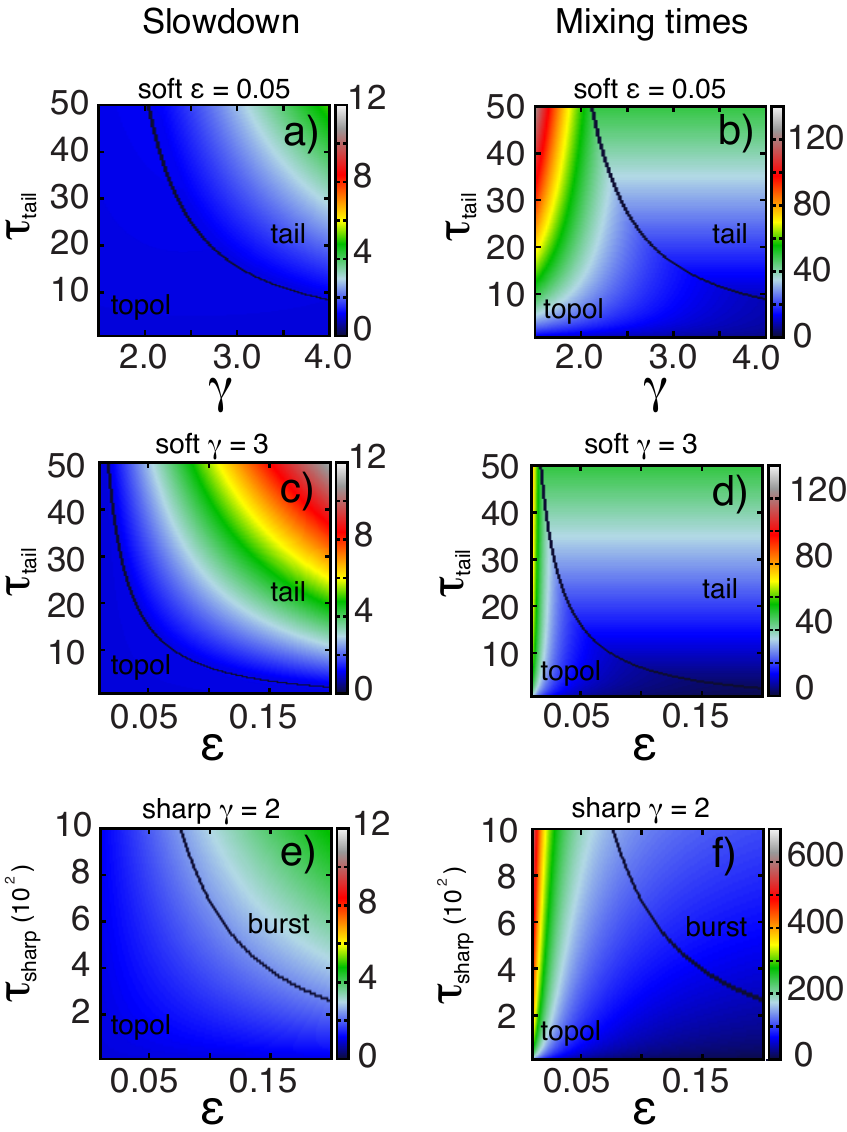}
\caption{\textbf{Slowdown factor and mixing times for power-law distributions of waiting times.} Panels (a,b) correspond to strong clustering (fixed $\epsilon=0.05$) and (c,d) to fixed $\gamma=3$ for soft cutoff; and (e,f) correspond to fixed $\gamma=2$ for sharp cutoff. Black curves divide the phases where either structure (topology), or burstiness, or tail dominates. The slowdown factor is the ratio between the mixing time of the power-law distribution of waiting times and of the exponential distribution (Poisson process). The value of the mixing time is taken as $\max(\mu(\epsilon^{-1}+\beta), \tau_{\textrm{tail}})$, always within $5 \%$ of the exact value as computed with equation (18) of the main text.}
\label{fig:SI}       
\end{figure}

As presented in the main text, in the extreme case of waiting times following a pure power-law, i.e.\ $\tau_{\textrm{tail}}=\infty$, infinite mixing time occurs irrespective of the network structure, as convergence to stationary state takes place at subexponential rate. This can be shown directly as after any time $t$, there will be a probability $1- \int_0^t \rho(r) dr$, polynomially decreasing, that the random walker has not moved at all from its initial node, making exponential convergence to stationarity impossible. If a soft (exponential) cutoff is added so that $\rho(t) \propto (t+1)^{-\gamma}\exp(-t/\tau_{\textrm{tail}})$, the slowdown is either driven by the tail or by the structure (Figure~\ref{fig:SI}b,d). Above a certain $\epsilon^*$, the tail completely determines the mixing time, which is not surprising considering that large $\epsilon$ corresponds to the absence of bottlenecks or strong clustering. Below this threshold, for instance at $\epsilon=0.05$ (Figure~\ref{fig:SI}a,b), there is a strong dependence on the structure for a range of exponents $\gamma$ and $\tau_{\textrm{tail}}$. In this scenario, burstiness never dominates, and the power-law exponent $\gamma$ does not play a direct role other than influencing $\mu$. Nevertheless, if the distribution of waiting times is modelled by a power-law with sharp cutoff, e.g.\ $\rho(t) \propto (t+1)^{-\gamma}$ for $0\leq t \leq \tau_{\textrm{sharp}}$, and $\rho(t) =0$ for $t>\tau_{\textrm{sharp}}$, there is no tail, and only burstiness and structure compete to regulate the slowdown (Figure~\ref{fig:SI}e,f).

These simple theoretical models allow to validate the accuracy of the approximation given by equation (12) in the main text for the mixing mode, $\tau_{\textrm{mix}} \approx \max(\mu(\epsilon^{-1}+\beta), \tau_{\textrm{tail}})$, to estimate the mixing time, as compared to the exact formula, see equation (18) in Methods B (error $<5\%$ on the examples of Figure~\ref{fig:SI}). They also allow to illustrate the fact that burstiness (dominating sharp cut-off power-laws) and fat tails (dominating soft cut-off power-laws) are independent features of a distribution of waiting times, as already exemplified in the main text.

\section*{SUPPLEMENTARY NOTE 2}
\label{II}
 
An empirical distribution of waiting times necessarily vanishes after some observation time $T$. Assume that the actual distribution is $\rho(\Delta t) = \alpha \rho_{0}(\Delta t) e^{-\Delta t/\tau_{\textrm{tail}}}$, where $\rho_{0}(\Delta t)$ is a fat tailed distribution and $\alpha$ a normalising constant. Assimilate the empirically observed distribution $\rho_{T}$ to the truncation $\rho(\Delta t)/\int_0^{T_{\textrm{obs}}} \rho$ until $\Delta t=T$ and zero for $\Delta t>T$. Then as $T$ grows to infinity, the moments of empirical and actual distribution  will coincide. As for the empirical Laplace transform $\rho_{T}(s)$, it is defined and analytic on the whole complex plane. We know that the Laplace transform of the actual distribution $\rho(\Delta t)$ cannot be extended to the left of $s=-1/\tau_{\textrm{tail}}$. What happens to $\rho_{T}(s)$ around those values? Choosing $s=-1/\tau_{\textrm{tail}}-\Delta s$, we write $\rho_T(s)=\int \rho_T(\Delta t) e^{-s\Delta t} d \Delta t$. If assume for simplicity  $\rho_{0}(\Delta t)$ to be decreasing, then a simple calculation yields $\rho_T(s) \geq \alpha \rho_{0}(T) (e^{\Delta s T}-1)/\Delta s$. Therefore $\rho_T(s)$, for $s$ even slightly below $-1/\tau_{\textrm{tail}}$, increases exponentially with $T$, thus is practically vertical for a high enough value of $T$. Therefore whenever a mode has characteristic time $\tau_k=\tau_{\textrm{tail}}$, the computation of $\tau_k$ from the empirical distribution and equation (17-18) in Methods B, $\rho_T(-1/\tau_k)=(1+\lambda_k)^{-1}$, is expected to be close to $\tau_k \approx \tau_{\textrm{tail}}$ as well. We deduce that provided a high enough observation time, the modelling of the distribution by a truncated distribution over the finite interval or by the extrapolation to a form $\rho_{0}(\Delta t) e^{-\Delta t/\tau_{\textrm{tail}}}$ is irrelevant, as both will lead to the same conclusions.

This fact leads to an interesting self-consistency condition as if the observation time is sufficiently long, we find that both the truncated model $\rho_T$ (via the equation $\rho_T(-1/\tau_{\textrm{mix}})=(1-\epsilon)^{-1}$) or the full-timeline model $\rho$ (leading to $\tau_{\textrm{mix}}=\tau_{\textrm{tail}}$ for fat-tail dominated networks) conclude to the same mixing time. If in addition, the approximation of equation (12) of the main text is valid for the mixing mode ($k=1$), we find the equation

\begin{equation}
\tau_{\textrm{mix}} \approx \mu(\epsilon^{-1}+ \beta) \approx \tau_{\textrm{tail}}, 
\label{eq:selfcons}
\end{equation}
indeed approximately verified by the  face-to-face networks dominated by fat tail effect (see Fig.~\ref{fig3arxiv}), especially SPM and SPH. A strong discrepancy between the sharp cut-off and full timeline model may reveal an excessively short observation time, insufficient for a reliable estimation of the mean, variance and/or tail of the distribution of waiting times, as is the case for SPC.

\section*{SUPPLEMENTARY NOTE 3}
\label{III}

The mathematical framework proposed in the main text adresses the combined effect of the network topology and the distribution of waiting times. Temporal networks obtained from real data however contain a number of temporal correlations and patterns, such as non-stationarity~\cite{Rocha13,Holme13,Horvath14} and causality~\cite{Scholtes2014}, or structure-time correlations~\cite{karsai2011small,Kivela12}. A careful analysis of the various patterns and correlations present in real data is out of the scope of our analysis. In order to contrast the limitations of our theory to the dynamics in real settings, we test our estimations by simulating a random walk dynamics directly on the empirical network such that the effect of all correlations and patterns of both time and structure are captured.

\begin{table}[h]
\centering
\begin{tabular}{ccccccccc}
\hline
    & $\delta t$   & $\mu$    & $\sigma$   & $\tau_{\textrm{tail}}$ & $\epsilon$ & $\beta$ & $\tau_{\textrm{mix}}$ & $M$\\
\hline
SPM & min   &   3.3   & 15.7   & 48.8    & 0.142 & 10.8 & 59.0   & Tail \\
SPC & min    & 16.2  & 101.2 & 182.5  & 0.073 & 19.0 & 529.8 & Burstiness \\
EMA & hour  & 14.5   & 59.4   & 177.4  & 0.074 & 7.9 & 310.4 & Structure \\
SEX & week & 18.0   & 32.2   & 20.2    & 0.012 & 1.1 & 1519.8  & Structure \\
\hline
\end{tabular}
\caption{\textbf{Summary statistics of the empirical networks.} Temporal resolution ($\delta t$);  mean ($\mu$) and standard deviation ($\sigma$) of the waiting time distribution; $\tau_{\textrm{tail}}$ is the least-square estimate of the tail characteristic time of the cumulative distribution of waiting times; $\epsilon$ is the spectral gap  for the aggregated weighted network; $\tau_{\textrm{mix}}$ is formally computed as $\max(\mu \epsilon^{-1}, \mu \beta, \tau_{\textrm{tail}})$ in analogy with the main text's case, even though this is not necessarily representative of the actual mixing time in this situation; and $M$ refers to the regulating mechanism that determines $\tau_{\textrm{mix}}$. We use the largest connected component.}
\label{tab_1_SI}
\end{table}

In the empirical temporal network each link has a time-stamp to indicate exactly the time it is active (being inactive at other times). The link to empirical temporal networks is done by performing a random walk on the time-stamped empirical data. We start a large number of walkers on each node at $t=0$. As time goes by, a walker remains in the node until one of its links becomes active and then jumps to the corresponding neighbour. The potential paths for the random walk are entirely defined by the empirical sequence of contacts (capturing temporal and temporal/structure correlations), meaning that not only the inter-contact time distribution is regulating the diffusion. This is in contrast to the experiments presented in the main text, where the dynamic networks had the  empirical structure (unweighted) fixed and the activation times of nodes were sampled from the empirical inter-contact time distribution (this process generates uncorrelated activation times following the empirical inter-contact time distribution). The spectral gap is calculated using the weighted version of the network, therefore inhomogeneities in the weight-topology are captured. The distribution of waiting times of the walkers provides us with empirical measures of $\mu$, $\sigma^2$ and $\tau_{\textrm{tail}}$. We apply this process to four of the datasets studied in the main text: face-to-face interactions between visitors in a museum (SPM) and between conference attendees (SPC)~\cite{Isella11}, to email communication within a university (EMA)~\cite{Eckmann04} and to sexual contacts between sex-workers and -buyers (SEX)~\cite{Rocha10} (see Table~\ref{tab_1_SI}). We underline that the two experiments, from this section and from Fig.~\ref{fig3arxiv}, use different methodologies, as the latter removed correlations or non-stationarity, but also weights on the network (allowing  to perform structure randomisation in Fig.~\ref{fig3arxiv}b,d -- main text -- conveniently). Therefore they are not supposed to deliver the same temporal or structural characteristics, except in order of magnitude. Moreover the mixing time in the diffusion process involving all correlations is not necessarily given by the formula $\tau_{\textrm{mix}} \approx \max(\mu \epsilon^{-1}, \mu \beta, \tau_{\textrm{tail}})$, which should be updated to account for event-event correlations, daily or weekly periodicity, etc. Nevertheless among those three mechanisms included in the formula, we notice a domination of waiting time effects, tail or burstiness in the SPC and SPM datasets. On the other hand, structural bottlenecks seem to be strong enough to dominate waiting times effects for EMA and SEX datasets, consistently with the experiment results in Fig.~\ref{fig3arxiv} in the main text. In future work, it would be interesting to check if supplementary correlations may reverse the respective strength of structure and waiting time effects.

\end{footnotesize}


\begin{thebibliography}{63}
\providecommand{\natexlab}[1]{#1}
\providecommand{\url}[1]{\texttt{#1}}
\expandafter\ifx\csname urlstyle\endcsname\relax
  \providecommand{\doi}[1]{doi: #1}\else
  \providecommand{\doi}{doi: \begingroup \urlstyle{rm}\Url}\fi

\bibitem[Bansal et~al.(2010)Bansal, Read, Pourbohloul, and Meyers]{Bansal10}
Bansal, S., Read, J., Pourbohloul, B. \& Meyers, L.~A.
\newblock The dynamic nature of contact networks in infectious disease epidemiology.
\newblock \emph{J. Biol. Dynam.} { \bf 4}, 478--489 (2010).

\bibitem[Barrat et~al.(2012)Barrat, Barth\'elemy, and Vespignani]{Barrat12}
Barrat, A., Barth\'elemy, M. \& Vespignani, A.
\newblock \emph{Dynamical {P}rocesses on {C}omplex {N}etworks.}
\newblock Cambridge University Press, Cambridge (2012).

\bibitem[Moody(2009)]{Moody09}
Moody, J.
\newblock \emph{The {O}xford {H}andbook of {A}nalytical {S}ociology.} chapter Network {D}ynamics., pages 447--–74.
\newblock Oxford University Press, Oxford (2009).

\bibitem[Newman(2010)]{Newman10}
Newman, M.E.J.
\newblock \emph{Networks: {A}n {I}ntroduction.}
\newblock Oxford University Press, Oxford (2010).

\bibitem[Vespignani(2012)]{Vespignani12}
Vespignani, A.
\newblock Modelling dynamical processes in complex socio-technical systems.
\newblock \emph{Nature Phys.} { \bf 8}, 32--39 (2012).

\bibitem[Chung(1996)]{Chung96}
Chung, F.R.K.
\newblock \emph{Spectral {G}raph {T}heory.}
\newblock American Mathematical Society (1996).

\bibitem[Lov\'asz(1993)]{Lovasz1993}
Lov\'asz, L.
\newblock Random walks on graphs: {A} survey.
\newblock \emph{Boy. Soc. Math. Stud.} { \bf 2}, 1--46 (1993).

\bibitem[Holme and Saram\"aki(2012)]{Holme12}
Holme, P. \& Saram\"aki, J.
\newblock Temporal networks.
\newblock \emph{Phys. Rep.} { \bf 519}, 97--125 (2012).

\bibitem[Barab\'asi(2005)]{Barabasi05}
Barab\'asi, A.-L.
\newblock The origin of bursts and heavy tails in human dynamics.
\newblock \emph{Nature} { \bf 435}, 207--211 (2005).

\bibitem[Eckmann et~al.(2004)Eckmann, Moses, and Sergi]{Eckmann04}
Eckmann, J.-P., Moses, E. \& Sergi, D.
\newblock Entropy of dialogues creates coherent structures in e-mail traffic.
\newblock \emph{Proc. Nat. Acad. Sci.} { \bf 101}, 14333--14337 (2004).

\bibitem[Haerter et~al.(2012)Haerter, Jamtveit, and Mathiesen]{Haerter12}
Haerter, J.O., Jamtveit, B. \& Mathiesen, J.
\newblock Communication dynamics in finite capacity social networks.
\newblock \emph{Phys. Rev. Lett.} { \bf 109}, 168701 (2012).

\bibitem[Holme et~al.(2004)Holme, Edling, and Liljeros]{Holme04}
Holme, P., Edling, C.R. \& Liljeros, F.
\newblock Structure and time-evolution of an {I}nternet dating community.
\newblock \emph{Soc. Net.} { \bf 26}, 2, 155--174 (2004).

\bibitem[Isella and et~al.(2011)]{Isella11}
Isella, L. et~al.
\newblock What's in a crowd? {A}nalysis of face-to-face behavioral networks.
\newblock \emph{J. Theo. Biol.} { \bf 271}, 1, 166--180 (2011).

\bibitem[Rocha et~al.(2010)Rocha, Liljeros, and Holme]{Rocha10}
Rocha, L.E.C., Liljeros, F. \& Holme, P.
\newblock Information dynamics shape the sexual networks of {I}nternet-mediated prostitution.
\newblock \emph{Proc. Nat. Acad. Sci.} { \bf 107}, 13, 5706--5711 (2010).

\bibitem[Starnini et~al.(2012)Starnini, Baronchelli, Barrat, and
  Pastor-Satorras]{Starnini12}
Starnini, M., Baronchelli, A., Barrat, A. \& Pastor-Satorras, R.
\newblock Random walks on temporal networks.
\newblock \emph{Phys. Rev. E} { \bf 85}, 5, 056115 (2012).

\bibitem[Vanhems and et~al.(2013)]{Vanhems13}
Vanhems, P. et~al.
\newblock Estimating potential infection transmission routes in hospital wards
  using wearable proximity sensors.
\newblock \emph{PLOS ONE} { \bf 8}, 9, e73970 (2013).

\bibitem[Rosvall et~al.(2014)Rosvall, Esquivel, Lancichinetti, West, and Lambiotte]{Rosvall2014}
Rosvall, M., Esquivel, A.V., Lancichinetti, A., West, J.D. \& Lambiotte, R.
\newblock Memory in network flows and its effects on spreading dynamics and community detection.
\newblock \emph{Nature Comm.} { \bf 5}, 4630 (2014).

\bibitem[Scholtes(2014)]{Scholtes2014}
Scholtes, I. et~al.
\newblock Causality-driven slow-down and speed-up of diffusion in non-{M}arkovian temporal networks.
\newblock \emph{Nature Comm.} { \bf 5}, 5024 (2014).

\bibitem[Holme and Liljeros(2014)]{Holme13}
Holme, P. \& Liljeros, F.
\newblock Birth and death of links control disease spreading in empirical contact networks.
\newblock \emph{Sci. Rep.} { \bf 4}, 4999 (2014).

\bibitem[Rocha and Blondel(2013)]{Rocha13}
Rocha, L.E.C. \& Blondel, V.~D.
\newblock Bursts of vertex activation and epidemics in evolving networks.
\newblock \emph{PLoS Comput. Biol.} { \bf 9}, 3, e1002974 (2013).

\bibitem[Horv\'ath and Kert\'esz(2014)]{Horvath14}
Horv\'ath, D.X. \& Kert\'esz, J.
\newblock Spreading dynamics on networks: the role of burstiness, topology and non-stationarity.
\newblock \emph{New J. Phys.} { \bf 16}, 7, 073037 (2014).

\bibitem[Malmgren et~al.(2008)Malmgren, Stouffer, Motter, and Amaral]{Malmgren08}
Malmgren, R.D., Stouffer, D.B., Motter, A.E. \& Amaral, L.A.N.
\newblock A {P}oissonian explanation for heavy tails in e-mail communication.
\newblock \emph{Proc. Nat. Acad. Sci.} { \bf 105}, 47, 18153--18158 (2008).

\bibitem[Perra and et~al.(2012)]{Perra2012}
Perra, N. et~al.
\newblock Random walks and search in time-varying networks.
\newblock \emph{Phys. Rev. Lett.} { \bf 109}, 238701 (2012).

\bibitem[Klafter and Sokolov(2011)]{Klafter11}
Klafter, J. \& Sokolov, I.M.
\newblock \emph{First {S}teps in {R}andom {W}alks: {F}rom {T}ools to {A}pplications.}
\newblock Oxford University Press, Oxford (2011).

\bibitem[Iribarren and Moro(2009)]{Iribarren09}
Iribarren, J.L. \& Moro, E.
\newblock Impact of human activity patterns on the dynamics of information  diffusion.
\newblock \emph{Phys Rev. Lett.} { \bf 103}, 3, 038702 (2009).

\bibitem[Jo et~al.(2014)Jo, Perotti, Kaski, and Kert\'esz]{Jo14}
Jo, H.-H., Perotti, J.I., Kaski, K. \& Kert\'esz, J.
\newblock Analytically solvable model of spreading dynamics with non-{P}oissonian.
\newblock \emph{Phys. Rev. X} { \bf 4}, 011041 (2014).

\bibitem[Min et~al.(2011)Min, Goh, and Vazquez]{Min11}
Min, B., Goh, K.-I. \& Vazquez, A.
\newblock Spreading dynamics following bursty human activity patterns.
\newblock \emph{Phys. Rev. E} { \bf 83}, 3, 036102 (2011).

\bibitem[Vazquez et~al.(2007)Vazquez, R\'acz, Luk\'acs, and
  Barab\'asi]{Vazquez07}
Vazquez, A., R\'acz, B., Luk\'acs, A. \& Barab\'asi, A.-L.
\newblock Impact of non-{P}oissonian activity patterns on spreading processes.
\newblock \emph{Phys. Rev. Lett.} { \bf 98}, 15, 158702 (2007).

\bibitem[Astr{\"o}m and Murray(2010)]{astrom2010feedback}
Astr{\"o}m, K.J. \& Murray, R.M.
\newblock \emph{Feedback {S}ystems: {A}n {I}ntroduction for {S}cientists and {E}ngineers.}
\newblock Princeton University Press, Princeton (2010).

\bibitem[Fax and Murray(2004)]{Fax04}
Fax, J.A. \& Murray, R.M.
\newblock Information flow and cooperative control of vehicle formations.
\newblock \emph{Autom. Cont.} { \bf 49}, 9, 1465--1476 (2004).

\bibitem[Blondel et~al.(2005)Blondel, Hendrickx, Olshevsky, and
  J.N.]{Blondel2005}
Blondel, V.D., Hendrickx, J.M., Olshevsky, A. \& J.N., Tsitsiklis.
\newblock Convergence in multiagent coordination, consensus, and flocking.
\newblock \emph{Proc. 44th IEEE Conf. Decision Control} { \bf }, 2996--3000 (2005).

\bibitem[Jadbabaie et~al.(2003)Jadbabaie, Lin, and Morse]{ali}
Jadbabaie, A., Lin, J. \& Morse, A.S.
\newblock Coordination of groups of mobile autonomous agents using nearest
  neighbor rules.
\newblock \emph{Autom. Control, IEEE Trans.} { \bf 48}, 6, 988--1001 (2003).

\bibitem[D\"{o}rfler and Bullo(2012)]{dorfler2012synchronization}
D\"{o}rfler, F. \& Bullo, F.
\newblock Synchronization and transient stability in power networks and nonuniform {K}uramoto oscillators.
\newblock \emph{SIAM J. Cont. Optim.} { \bf 50}, 3, 1616--1642 (2012).

\bibitem[Strogatz(2000)]{Strogatz2000}
Strogatz, S.H.
\newblock From {K}uramoto to {C}rawford: {E}xploring the onset of synchronization of in populations of coupled oscillators.
\newblock \emph{Phys. D} { \bf 143}, 1-4, 1--20 (2000).

\bibitem[Yin & Zhang(2012)]{yin2012continuous}
Yin, G.G. \& Zhang, Q.
\newblock \emph{Continuous-time {M}arkov {C}hains and {A}pplications: {A} {T}wo-time-scale {A}pproach.} volume~37.
\newblock Springer (2012).

\bibitem[Levin et~al.(2008)Levin, Peres, and Wilmer]{Levinmixing}
Levin, D.A., Peres, Y. \& Wilmer, E.L.
\newblock \emph{{M}arkov {C}hains and {M}ixing {T}imes.}
\newblock American Mathematical Society (2008).

\bibitem[Hoffmann et~al.(2012)Hoffmann, Porter, and Lambiotte]{Hoffmann12}
Hoffmann, T., Porter, M.A. \& Lambiotte, R.
\newblock Generalized master equations for non-{P}oisson dynamics on networks.
\newblock \emph{Phys. Rev. E} { \bf 86}, 4, 046102 (2012).

\bibitem[Montroll and Weiss(1965)]{montroll1965random}
Montroll, E.W. \& Weiss, G.H.
\newblock Random walks on lattices ii.
\newblock \emph{J. Math. Phys.} { \bf 6}, 2, 167--181 (1965).

\bibitem[Kolmogoroff(1931)]{kolmogoroff1931analytischen}
Kolmogoroff, A.
\newblock {U}ber die analytischen {M}ethoden in der {W}ahrscheinlichkeitsrechnung.
\newblock \emph{Math. Ann.} { \bf 104}, 1, 415--458 (1931).

\bibitem[Dyke(2014)]{dyke1999introduction}
Dyke, P.
\newblock \emph{An {I}ntroduction to {L}aplace {T}ransforms and {F}ourier {S}eries.}
\newblock Springer (2014).

\bibitem[Keeling and Ross(2008)]{keeling2008methods}
Keeling, M.J. \& Ross, J.V.
\newblock On methods for studying stochastic disease dynamics.
\newblock \emph{J. R. Soc. Interface} { \bf 5}, 19, 171--181 (2008).

\bibitem[Kivel\"a(2012)]{Kivela12}
Kivel\"a, M. et~al.
\newblock Multiscale analysis of spreading in a large communication network.
\newblock \emph{J. Stat. Mech.} { \bf 2012}, 03, P03005 (2012).

\bibitem[Rocha et~al.(2011)Rocha, Liljeros, and Holme]{Rocha11}
Rocha, L.E.C., Liljeros, F. \& Holme, P.
\newblock Simulated epidemics in an empirical spatiotemporal network of 50,185 sexual contacts.
\newblock \emph{PLoS Comput. Biol.} { \bf 7}, 3, e1001109 (2011).

\bibitem[Speidel et~al.(2015)Speidel, Lambiotte, Aihara, and Masuda]{Speidel}
Speidel, L., Lambiotte, R., Aihara, K. \& Masuda, N.
\newblock Steady state and mean recurrence time for random walks on stochastic temporal networks.
\newblock \emph{Phys. Rev. E} { \bf 91}, 1, 012806 (2015).

\bibitem[Kleinrock(1975)]{kleinrock1975queueing}
Kleinrock, L.
\newblock Queueing systems: {T}heory. {I}.
\newblock { \bf } (1975).

\bibitem[Lambiotte et~al.(2013)Lambiotte, Tabourier, and Delvenne]{Lambiotte13}
Lambiotte, R., Tabourier, L. \& Delvenne, J.-C.
\newblock Burstiness and spreading on temporal networks.
\newblock \emph{Eur. Phys. J. B} { \bf 86}, 7 (2013).

\bibitem[Karsai(2011)]{karsai2011small}
Karsai, M. et~al.
\newblock Small but slow world: {H}ow network topology and burstiness slow down spreading.
\newblock \emph{Phys. Rev. E} { \bf 83}, 2, 025102 (2011).

\bibitem[Goh and Barab\'asi(2008)]{Goh08}
Goh, K.I. \& Barab\'asi, A.-L.
\newblock Burstiness and memory in complex systems.
\newblock \emph{Europhys. Lett.} { \bf 81}, 4 (2008).

\bibitem[Cheeger(1970)]{Cheeger70}
Cheeger, J.
\newblock \emph{Problems in {A}nalysis.} chapter A lower bound for the smallest eigenvalue of the {L}aplacian., pages 195--199.
\newblock Princeton University Press, Princeton, N.J. (1970).

\bibitem[Diaconis and Stroock(1991)]{Diaconis91}
Diaconis, P. \& Stroock, D.
\newblock Geometric bounds for eigenvalues of {M}arkov chains.
\newblock \emph{Ann. Appl. Probab.} { \bf 1}, 1, 36--61 (1991).

\bibitem[Gfeller and De~Los~Rios(2008)]{Gfeller}
Gfeller, D. \& De~Los~Rios, P.
\newblock Spectral coarse graining and synchronization in oscillator networks.
\newblock \emph{Phys. Rev. Lett.} { \bf 100}, 17, 174104 (2008).

\bibitem[Kokotovic et~al.(1987)Kokotovic, Khalil, and
  O'reilly]{koko1999singular}
Kokotovic, P., Khalil, H.K. \& O'reilly, J.
\newblock \emph{Singular {P}erturbation {M}ethods in {C}ontrol: {A}nalysis and {D}esign.} volume~25.
\newblock Society for Industrial and Applied Mathematics (1987).

\bibitem[Simon and Ando(1961)]{simon1961aggregation}
Simon, H.A. \& Ando, A.
\newblock Aggregation of variables in dynamic systems.
\newblock \emph{Econometrica} { \bf }, 111--138 (1961).

\bibitem[Simonsen(2005)]{simonsen2005diffusion}
Simonsen, I.
\newblock Diffusion and networks: {A} powerful combination!
\newblock \emph{Phys. A} { \bf 357}, 2, 317--330 (2005).

\bibitem[Delvenne et~al.(2013)Delvenne, Schaub, Yaliraki, and
  Barahona]{Delvenne13}
Delvenne, J.-C., Schaub, M.T., Yaliraki, S.N. \& Barahona, M.
\newblock \emph{Dynamics {O}n and {O}f {C}omplex {N}etworks.} volume~2, chapter The stability of a graph partition: {A} dynamics-based framework for community detection., pages 221--242.
\newblock Springer (2013).

\bibitem[Shen and Cheng(2010)]{shen2010spectral}
Shen, H.-W. \& Cheng, X.-Q.
\newblock Spectral methods for the detection of network community structure: {A} comparative analysis.
\newblock \emph{J. Stat. Mech.} { \bf 2010}, 10, P10020 (2010).

\bibitem[Von~Luxburg(2007)]{vonluxburg2007}
Von~Luxburg, U.
\newblock A tutorial on spectral clustering.
\newblock \emph{Stat. Comput.} { \bf 17}, 4, 395--416 (2007).

\bibitem[Reichardt and Bornholdt(2006)]{reichardt}
Reichardt, J. \& Bornholdt, S.
\newblock Statistical mechanics of community detection.
\newblock \emph{Phys. Rev. E} { \bf 74}, 016110 (2006).

\bibitem[Ronhovde and Nussinov(2009)]{ronhovde}
Ronhovde, P. \& Nussinov, Z.
\newblock Multiresolution community detection for megascale networks by information-based replica correlations.
\newblock \emph{Phys. Rev. E} { \bf 80}, 1, 016109 (2009).

\bibitem[Fortunato(2010)]{Fortunato10}
Fortunato, S.
\newblock Community detection in graphs.
\newblock \emph{Phys. Rep.} { \bf 486}, 75--174 (2010).

\bibitem[Karrer et~al.(2008)Karrer, Levina, and Newman]{karrer}
Karrer, B., Levina, E. \& Newman, M.E.J.
\newblock Robustness of community structure in networks.
\newblock \emph{Phys. Rev. E} { \bf 77}, 4, 046119 (2008).

\bibitem[Delmotte et~al.(2011)Delmotte, Tate, Yaliraki, and
  Barahona]{Delmotte2011}
Delmotte, A., Tate, E.W., Yaliraki, S.N. \& Barahona, M.
\newblock Protein multi-scale organization through graph partitioning and
  robustness analysis: {A}pplication to the myosin light chain interaction.
\newblock \emph{Phys. Biol.} { \bf 8}, 5, 055010 (2011).


\bibitem[Delmotte et~al(2012)Delmotte, Schaub, Yaliraki, and Barahona]{StabilityToolbox}
\newblock Delmotte, A., Schaub, M.T., Yaliraki, S.N. \& Barahona, M., Community Detection using the stability of a graph partition. \newblock https://github.com/michaelschaub/PartitionStability (2012).


%%%%%%%%%%%%%%



\bibitem{miyazako2013turing}
H.~Miyazako, Y.~Hori, S.~Hara Turing instability in reaction-diffusion systems
  with a single diffuser: {C}haracterization based on root locus. in: Dec.
  Cont. (CDC), IEEE 52nd Ann. Conf. 2671--2676 (2013).

\bibitem{brin1998anatomy}
S.~Brin, L.~Page The anatomy of a large-scale hypertextual {W}eb search
  engine. Comp. Net. ISDN Syst. 30~(1) 107--117 (1998).



\bibitem{rosvall2008maps}
M.~Rosvall, C.~T. Bergstrom Maps of random walks on complex networks reveal
  community structure., Proc. Nat. Acad. Sci. 105~(4) 1118--1123 (2008).



\bibitem{hethcote2000mathematics}
H.~W. Hethcote The mathematics of infectious diseases. SIAM Rev. 42~(4) 599--653 (2000).




\bibitem{menck2013basin}
P.~J. Menck, J.~Heitzig, N.~Marwan, J.~Kurths How basin stability complements
  the linear-stability paradigm. Nature Phys. 9~(2) 89--92 (2013).



\end{thebibliography}
\end{document}